\documentclass[twocolumn,amssymb,secnumarabic,preprintnumbers, amsmath]{revtex4}
\pdfoutput=1

\usepackage{graphicx}
\usepackage{booktabs}
\usepackage{placeins}

\newcommand{\C}[1]{{\cal C}_{#1}}

\newcommand{\av}[1]{\langle #1 \rangle}

\begin{document}

\title{Patterns of New Physics in $b\to s \ell^+\ell^-$ transitions in the light of recent data}

\author{Bernat Capdevila$^{a}$, Andreas Crivellin$^{b}$, S\'ebastien Descotes-Genon$^{c}$, Joaquim Matias$^{*a}$ and Javier
Virto$^{d}$
\vspace{0.3cm}}

\affiliation{
\mbox{$^{a}$Universitat Aut\`onoma de Barcelona, 08193 Bellaterra, Barcelona,
Institut de Fisica d'Altes Energies (IFAE),}\\ 
\mbox{The Barcelona Institute of Science and Technology, Campus UAB, 08193 Bellaterra (Barcelona), Spain}\\
\mbox{$^{b}$Paul Scherrer Institut, CH--5232 Villigen PSI, Switzerland}\\
\mbox{$^{c}$Laboratoire de Physique Th\'eorique, UMR 8627,}\\ 
\mbox{CNRS, Univ. Paris-Sud, Universit\'e Paris-Saclay, 91405 Orsay Cedex, France}\\
\mbox{$^{d}$Albert Einstein Center for Fundamental Physics, Institute for Theoretical Physics,}\\ 
\mbox{University of Bern, CH-3012 Bern, Switzerland}\\
}



\begin{abstract}
In the Standard Model (SM), the rare transitions where a bottom quark decays into a strange quark and a pair of light leptons 
 exhibit a potential sensitivity to physics beyond the SM. In addition, the SM embeds Lepton Flavour Universality (LFU), leading to almost identical probabilities for muon and electron modes. The LHCb collaboration discovered a set of deviations from the SM expectations in decays to muons and also in ratios assessing LFU.  Other experiments (Belle, ATLAS, CMS) found consistent measurements, albeit with large error bars. We perform a global fit to all available $b\to s\ell^+\ell^-$ data ($\ell=e,\mu$) in a model-independent way allowing for different patterns of New Physics. For the first time, the NP hypothesis is preferred over the SM by $5\,\sigma$ in a general case when NP can enter  SM-like operators and their chirally-flipped partners. LFU violation is favoured with respect to LFU at the 3-4$\,\sigma$ level. We discuss the impact of LFU-violating New Physics on the observable $P_5^\prime$ from $B \to K^*\mu^+\mu^-$ and we compare our estimate for  long-distance charm contributions with an empirical model recently proposed by a group of LHCb experimentalists. Finally, we discuss NP models able to describe this consistent pattern of deviations. 
\end{abstract}

\pacs{13.25.Hw, 11.30.Er, 11.30.Hv}

\maketitle

\begin{table*}[t]
\begin{tabular}{c||c|c|c|c|c|c|c} 
Largest pulls & $\av{P_5^\prime}_{[4,6]}$ & $\av{P_5^\prime}_{[6,8]}$ & $R_K^{[1,6]}$ & $R_{K^*}^{[0.045,1.1]}$ & $R_{K^*}^{[1.1,6]}$ & ${\cal B}_{B_s \to \phi\mu^+\mu^-}^{[2,5]}$  &  ${\cal B}_{B_s \to \phi\mu^+\mu^-}^{[5,8]}$  \\
\hline\hline
Experiment & $-0.30 \pm 0.16$ & $-0.51 \pm 0.12$  & $0.745^{+0.097}_{-0.082}$  & $0.66^{+0.113}_{-0.074}$ & $0.685^{+0.122}_{-0.083}$  & $ 0.77 \pm 0.14$ & $0.96 \pm 0.15$ \\ 
\hline
SM prediction & $-0.82 \pm 0.08$  & $-0.94 \pm 0.08$ & $1.00 \pm 0.01$  &$0.92 \pm 0.02$ & $1.00 \pm 0.01$  & $1.55 \pm 0.33$  & $1.88 \pm 0.39$  \\
Pull ($\sigma$)  & -2.9 & -2.9 & +2.6 & +2.3 & +2.6 & +2.2 & +2.2 \\
\hline
Prediction for $\C{9\mu}^{\rm NP}=-1.1$ & $-0.50\pm 0.11$ & $-0.73 \pm 0.12 $ & $0.79 \pm 0.01$ & $0.90 \pm 0.05$ & $0.87 \pm 0.08 $ & $1.30 \pm 0.26$ & $1.51 \pm 0.30 $\\
Pull ($\sigma$) & -1.0 & -1.3 &  +0.4 & +1.9  & +1.2  & +1.8  & +1.6
\end{tabular}
\caption{Main anomalies currently observed in $b\to s\ell\ell$ transitions, with the current measurements, our predictions for the SM and the NP scenario $\C{9\mu}^{\rm NP}=-1.1$, and the corresponding pulls. In addition, a deficit compared to the SM predictions has been observed at low and large recoils for ${\cal B}(B^{(0,+)} \to K^{(0,+)} \mu\mu)$~\cite{Aaij:2014pli} and ${\cal B}(B^0 \to K^{*0} \mu\mu)$~\cite{Aaij:2016flj}, as well as at low recoil (above 15 GeV$^2$) for ${\cal B}(B^+ \to K^{*+}\mu^+\mu^-)$~\cite{Aaij:2014pli} and ${\cal B}(B_s \to \phi \mu^+\mu^-)$~\cite{Aaij:2015esa}.} \label{tensions}
\end{table*}

\section{Introduction}

The discovery of the Higgs boson marked the completion of the Standard Model (SM) of particle physics, which describes the elementary constituents of matter and their interactions (strong, electromagnetic and weak) as a quantum field theory. This has led to precise predictions for measurable quantities tested experimentally with a high accuracy. Now, the main focus has shifted to the identification of the physics beyond the SM occurring at higher energies. While the LHC has not observed new heavy particles directly, indirect searches having access to higher scales through flavour observables have evolved from a precision study  towards a search tool for New Physics (NP).

Over the last few years, many observables related to the flavour-changing neutral-current transitions $b\to s\ell^+\ell^-$ have exhibited deviations from SM expectations. Due to their suppression within the SM, these transitions are well known to have a high sensitivity to potential NP contributions. In order to evaluate the significance and coherence of these deviations, a global model-independent fit is the most efficient tool to determine if they contain patterns explained by NP in a consistent way.

The present situation is  {\it exceptional} in the sense that we have found that the observed deviations  indeed form coherent patterns within the model-independent approach of the effective Hamiltonian governing the $b \to s \ell\ell$ transitions. Already in 2013  
first hints of this consistency were pointed out in Ref. \cite{Descotes-Genon:2013wba}  (using only $B\to K^*\mu\mu$) and later on in Ref. \cite{Descotes-Genon:2015uva} (with all LHCb data available at that time)   showing that a very economical mechanism, namely a negative contribution of the order of $-25\%$  to the short-distance coefficient of the effective operator ${\cal O}_{9\mu}=\frac{e^2}{16 \pi^2} \left( \bar{s} \gamma_\mu P_L b \right)\left(\bar{\mu}\gamma^\mu\mu\right)$, is sufficient to alleviate all above-mentioned tensions, whereas the data allowed for NP contributions to other operators. 
This picture was later confirmed by other global analyses~\cite{Altmannshofer:2014rta,Hurth:2016fbr}  using different observables, hadronic inputs and theory approaches for their computations. Recent experiment results have shown additional hints of NP, indicating a violation of Lepton Flavour Universality (LFU) between $b\to s ee$ and $b\to s \mu\mu$ processes. The purpose of the present article is to reconsider our global analysis including LFU-violating  (LFUV) ratios together with updated experimental and theoretical information on $b\to s \ell\ell$ processes. 

After a brief summary of the experimental situation in Sec.~\ref{sec:exp}, we recall the general framework of our model-independent analysis, i.e., the effective Hamiltonian in Sec.~\ref{sec:framework}. In Sec.~\ref{sec:results}, we probe 
various NP patterns through a global fit to the data, as for instance the scenario identified in Ref.~\cite{Descotes-Genon:2013wba} with New Physics in the short-distance Wilson coefficient $C_{9\mu}$ only, the
scenario $C_{9\mu}^{\rm NP}=-C_{10\mu}^{\rm NP}$ which is a very appealing one from the model building point of view,  and scenarios including operators with right-handed couplings to quarks (LFUV observables had already an important impact on models with right-handed currents~\cite{Becirevic:2015asa,Becirevic:2016yqi}). We discuss also alternative patterns
with NP affecting both muons and electrons, although not in the same way. In all these cases, we identify patterns of NP that are significantly preferred (above the $5\,\sigma$ level) compared to the SM and solve the deviations discussed at the beginning of this section. We also discuss the general scenario
where we allow for NP in all SM and chirally-flipped operators. 
In Sec.~\ref{sec:RKRD}, we discuss the role played by the LFUV observables $R_K$ and $R_{K^*}$ in our global analysis thanks to semi-analytical formulae.
In Sec.~\ref{sec:models}, we discuss the impact of these compelling patterns of physics beyond the SM on specific NP models with new heavy gauge bosons or leptoquarks (hypothetical particles coupling to quarks and leptons simultaneously), which provide extensions of the SM able to explain the anomalies.

We relate the impact of LFUV observables with the most prominent deviation among $b \to s \mu\mu$ observables, namely $P_5^\prime$~\cite{DescotesGenon:2012zf}, in Sec.~\ref{sec:hadtest}. We also provide additional information concerning theoretical uncertainties, focusing on the issue of charm-loop contributions: we perform a comparison between the empirical model analysed in Ref.~\cite{Blake:2017fyh}, and our estimate 
 based on the framework of Ref.~\cite{Khodjamirian:2010vf}, finding a very good agreement between the two.
In Sec.~\ref{sec:oppLFUV}, we investigate how additional LFUV observables can disentangle the various scenarios favoured by our global analyses in the future, before reaching our conclusions.

\section{Experimental situation}\label{sec:exp}

We start by briefly discussing the recent experimental activity concerning $b\to s\ell\ell$ transitions.
In 2013, using the 1~fb$^{-1}$ dataset, the LHCb experiment measured the basis of optimised observables~\cite{Descotes-Genon:2013vna} for $B\to K^*\mu^+\mu^-$~\cite{Aaij:2013qta}, observing the so-called $P_5^\prime$ anomaly~\cite{Descotes-Genon:2013wba}, i.e., a sizeable 3.7~$\sigma$ discrepancy between the measurement and the SM prediction in  one bin for the angular observable $P_5^\prime$~\cite{DescotesGenon:2012zf}. In 2015, using
the 3 fb~$^{-1}$ dataset, LHCb confirmed this discrepancy with a 3~$\sigma$ deviation in each of two adjacent bins at large $K^*$ recoil~\cite{Aaij:2015oid}. LHCb also observed a systematic deficit with respect to SM predictions for the branching ratios of several decays,~\cite{Aaij:2013aln,Aaij:2015esa}. In 2016, the Belle experiment presented an independent analysis of $P_5^\prime$~\cite{Abdesselam:2016llu,Wehle:2016yoi} confirming the LHCb measurements in a very different experimental setting.

A conceptually new element arose when a discrepancy
 in the ratio $R_K={\cal B}_{B \to K \mu^+\mu^-}$ $/{\cal B}_{B \to K e^+e^-}$ was also observed by LHCb~\cite{Aaij:2014ora}, hinting at the violation of Lepton Flavour Universality (LFU) and
suggesting that deviations from the SM are predominantly present in $b\to s\mu^+\mu^-$ transitions but not in $b\to s e^+e^-$ ones. Recently Belle has measured for the first time~\cite{Wehle:2016yoi} the additional LFU violating (LFUV) observables $Q_{4,5}=P_{4,5}^{\mu\prime}-P_{4,5}^{e\prime}$, proposed in Ref.~\cite{Capdevila:2016ivx}. Even if  not yet statistically significant, the result points also towards LFUV in $Q_5$, consistently with the deviation in $R_K$.

The ATLAS and CMS collaborations have presented new preliminary results for $B\to K^*\mu\mu$ observables: ATLAS measured the whole set as well as $F_L$ at large $K^*$ recoil~\cite{ATLAS:2017dlm}, whereas CMS presented results for $P_1$ and $P_5'$ at low and large recoils~\cite{CMS:2017ivg}. The results show a good (but not perfect) overall agreement with the LHCb results, and a global model-independent analysis~\cite{Altmannshofer:2017fio} has confirmed the earlier picture in Refs.~\cite{Descotes-Genon:2013wba,Descotes-Genon:2015uva,Altmannshofer:2014rta,Hurth:2016fbr} on many issues: favoured hypotheses for NP contributions to Wilson Coefficients, consistency of deviation patterns in the various channels and types of observables, robustness with respect to the theoretical assumptions on hadronic corrections, and absence of $q^2$- or helicity-dependences for $\C{9,\mu}^{\rm NP}$ that would signal uncontrolled long-distance contributions in $B\to K^*\mu^+\mu^-$.

On the other hand, the LHCb collaboration has recently updated the differential branching ratio for $B \to K^* \mu^+\mu^-$~\cite{Aaij:2016flj}, and it has presented striking new results concerning the LFUV ratio $R_{K^*}={\cal B}_{B \to K^* \mu^+\mu^-}$ $/{\cal B}_{B \to K^* e^+e^-}$ at large $K^*$ recoil~\cite{Aaij:2017vbb}, exhibiting significant deviations from SM expectations. Ratios like $R_K$ and $R_{K^*}$ are particularly interesting due to their lack of sensitivity to hadronic uncertainties in the SM and their potential to uncover NP~\cite{Hiller:2014yaa,Hiller:2014ula}.
The significant deviation of $R_{K^*}$ from SM expectations confirms in particular that hadronic uncertainties in the theoretical predictions are not sufficient to explain all the anomalies observed in $b\to s\ell^+\ell^-$ transitions, and that alternative explanations must be searched for.

A summary of the most prominent anomalies is presented in Table~\ref{tensions}.
In the following, we discuss how these remarkable new results affect the global model-independent analysis of NP in $b \to s \ell^+ \ell^-$ decays, we determine patterns of NP contributions favored by the whole set of experimental data, and discuss their implications for NP models as well as further experimental tests.

\section{General framework}\label{sec:framework}

In order to combine all measurements  and evaluate their impact, importance and consistency, one has to perform a global fit to all available data. We perform such a fit along the lines of Ref.~\cite{Descotes-Genon:2015uva}. Our starting point is an effective Hamiltonian~\cite{Grinstein:1987vj,Buchalla:1995vs} in which heavy degrees of freedom (the top quark, the $W$ and $Z$ bosons, the Higgs and any potential heavy new particles) have been integrated out in short-distance Wilson coefficients $\C{i}$, leaving only a set of operators $O_i$ describing the physics at long distances:
\begin{equation}
{\cal H}_{\rm eff}=-\frac{4G_F}{\sqrt{2}} V_{tb}V_{ts}^*\sum_i \C{i}  {\cal O}_i
\end{equation}
(up to small corrections proportional to $V_{ub}V_{us}^*$ in the SM). 
In the SM, the Hamiltonian contains 10 main operators with specific chiralities due to the $V-A$ structure of the weak interactions. In presence of NP, additional operators may become of importance. For the processes considered here, we focus our attention on the operators:
\begin{eqnarray}
{\mathcal{O}}_{7} &=& \frac{e}{16 \pi^2} m_b
(\bar{s} \sigma_{\mu \nu} P_R b) F^{\mu \nu} ,\\
{\mathcal{O}}_{{7}^\prime} &= &\frac{e}{16 \pi^2} m_b
(\bar{s} \sigma_{\mu \nu} P_L b) F^{\mu \nu} ,\\
{\mathcal{O}}_{9\ell} &=& \frac{e^2}{16 \pi^2} 
(\bar{s} \gamma_{\mu} P_L b)(\bar{\ell} \gamma^\mu \ell) ,\\
{\mathcal{O}}_{{9\ell}^\prime} &=& \frac{e^2}{16 \pi^2} 
(\bar{s} \gamma_{\mu} P_R b)(\bar{\ell} \gamma^\mu \ell) , \\
{\mathcal{O}}_{10\ell} &=&\frac{e^2}{16 \pi^2}
(\bar{s}  \gamma_{\mu} P_L b)(  \bar{\ell} \gamma^\mu \gamma_5 \ell) ,\\
{\mathcal{O}}_{{10\ell}^\prime} &=&\frac{e^2}{16\pi^2}
(\bar{s}  \gamma_{\mu} P_R b)(  \bar{\ell} \gamma^\mu \gamma_5 \ell) ,
\end{eqnarray}
where $P_{L,R}=(1 \mp \gamma_5)/2$ and $m_b \equiv m_b(\mu_b)$ denotes the running $b$ quark mass in the $\overline{\mathrm{MS}}$ scheme. 
Their associated Wilson coefficients are $\C7,\C{9\ell},\C{10\ell}$ and $\C7^\prime,\C{9\ell}^\prime,\C{10\ell}^\prime$ with $\ell = e$ or $\mu$.
$\C7^{(\prime)}$ describe the interaction strength of bottom ($b$) and strange ($s$) quarks with the photon while $\C{9\ell,10\ell}$ and $\C{9\ell,10\ell}^\prime$ encode the interaction strength of $b$ and $s$ quarks with charged leptons. $\C{9\ell,10\ell}$ and $\C{9\ell,10\ell}^\prime$ are equal for muons and electrons in the SM but NP can add very different contributions in muons compared to electrons. For $\C7$ and $\C{9\ell,10\ell}$ we split SM and NP contributions like ${\cal C}_{i\ell}={\cal C}_{i\ell}^{\rm SM}+ {\cal C}_{i\ell}^{\rm  NP}$ (the SM contributions to chirally-flipped operators are negligible). 

We include all the observables considered in the reference fit of Ref.~\cite{Descotes-Genon:2015uva} (see Secs.~2 and 3, and App.~A of this reference). More specifically, for the angular observables in $B\to K^\star\mu^+\mu^-$, $B\to K^\star e^+e^-$ and $B_s\to \phi\mu^+\mu^-$, we use the optimised observables $P_i^{(\prime)}$ obtained from LHCb's likelihood fit~\cite{Aaij:2015oid}. Concerning the $q^2$ binning we use the finest bins
at large recoil (below the $J/\psi$) but the widest bins in the low-recoil region to ensure quark-hadron duality. For the $b\to s\gamma$ radiative observables, we add to our previous set of observables the branching ratios of the radiative decays $B^0\to {K}^{*0}\gamma$, $B^+\to {K}^{*+}\gamma$, $B_s\to\phi\gamma$~\cite{Amhis:2016xyh}.

In addition, and following the discussion of the previous section, we add to the fit all the new measurements made available since Ref.~\cite{Descotes-Genon:2015uva}:\\[-2mm]

\noindent $\blacktriangleright$ The $B^0 \to K^{\star 0}\mu^+\mu^-$ differential branching fraction measured by LHCb~\cite{Aaij:2016flj} based on the full
run 1 dataset, superseding the results in Ref.~\cite{Aaij:2013iag}. We use the most  recent update of Ref.~\cite{Aaij:2016flj} that led to a reduction of the branching ratio by
about $20\%$ in magnitude.\\[-2mm]

\noindent $\blacktriangleright$ The new Belle measurements~\cite{Wehle:2016yoi}
for the isospin-averaged but lepton-flavour dependent $B \to K^{\star}\ell^+\ell^-$ observables $P_{4,5}^{\prime\,e}$ and $P_{4,5}^{\prime\,\mu}$. The isospin average
is given by the following expression~\cite{Wehle:private},
\begin{equation}
P_{i}^{\prime\,\ell} = \sigma_+\, P_{i}^{\prime\,\ell}(B^+)
+ (1-\sigma_+)\, P_{i}^{\prime\,\ell}(\bar B^0)\ .
\end{equation}
Since $\sigma_+$ describing the relative weight of each isospin component in the average is not public, we treat it as a nuisance parameter $\sigma_+ =0.5\pm 0.5$. This will not have a significant effect in our results, since the isospin breaking in the SM is small (but accounted for in our analysis), and we do not consider NP contributions to four-quark operators.\\[-2mm]

\noindent $\blacktriangleright$ The new ATLAS measurements~\cite{ATLAS:2017dlm} on the angular observables $P_1$,  $P'_{4,5,6,8}$ in $B^0 \to K^{\star 0}\mu^+\mu^-$ as well as $F_L$
in the large recoil region.\\[-2mm]

\noindent $\blacktriangleright$ The new CMS measurements~\cite{CMS:2017ivg} on the angular observables $P_1$ and $P'_5$ in $B^0 \to K^{\star 0}\mu^+\mu^-$, both at large and
low recoils (we consider only the bin at low recoil). We take $F_L$ and $A_{FB}$ from an earlier analysis~\cite{Khachatryan:2015isa}. We also include the data from an earlier analysis at 7 TeV~\cite{Chatrchyan:2013cda}. A very welcome check of the stability of the CMS results would consist in performing a simultaneous extraction of $F_L$, $P_1$ and $P_5'$, using the same folding distribution as ATLAS, LHCb and Belle.\\[-2mm]

\noindent $\blacktriangleright$ The new measurements of the lepton-flavour non-universality ratio $R_{K^\star}$ in two large-recoil bins by the LHCb collaboration~\cite{Aaij:2017vbb}. The likelihood of these measurements is asymmetric, and dominated by statistical uncertainties. We thus take the two measurements as uncorrelated, and for each of the two bins, we take a symmetric Gaussian error that is the larger of the two asymmetric uncertainties (while keeping the central value unchanged). This approach will underestimate the impact of these measurements on our fit, but we prefer to remain conservative on this point until the likelihood is known in detail.\\[-2mm]

Following Ref.~\cite{Descotes-Genon:2015uva}, we take into account the correlations whenever available, and assume that the measurements are uncorrelated otherwise. In order to avoid including measurements with too large correlations, we include the LHCb measurements of the ratios $R_{K^*}$ and $R_K$, as well as the differential branching ratios ${\cal B}(B^0\to K^{*0}\mu\mu)$ and ${\cal B}(B^+\to K^+\mu\mu)$,
but we discard ${\cal B}(B^0\to K^{*0}ee)^{[0.0009,1]}$ and ${\cal B}(B^+\to K^+ee)^{[1,6]}$.

\begin{table*}
\begin{tabular}{c||c|c|c|c|c||c|c|c|c|c}
 & \multicolumn{5}{c||}{All} &  \multicolumn{5}{c}{LFUV}\\
\hline
1D Hyp.   & Best fit& 1 $\sigma$ & 2 $\sigma$  & Pull$_{\rm SM}$ & p-value & Best fit & 1 $\sigma$ & 2 $\sigma$ & Pull$_{\rm SM}$ & p-value\\
\hline\hline
$\C{9\mu}^{\rm NP}$    & -1.11 &    $[-1.28,-0.94]$ & $[-1.45,-0.75]$   & 5.8   &  68  &
 -1.76    &   $[-2.36,-1.23]$ & $[-3.04,-0.76]$   & 3.9 & 69  \\
$\C{9\mu}^{\rm NP}=-\C{10\mu}^{\rm NP}$    & -0.62  &    $[-0.75,-0.49]$ & $[-0.88,-0.37]$   &  5.3  & 58  &   -0.66  &   $[-0.84,-0.48]$ & $[-1.04,-0.32]$   & 4.1  & 78 \\
$\C{9\mu}^{\rm NP}=-\C{9\mu}^{\prime}$     & -1.01 &    $[-1.18,-0.84]$ & $[-1.34,-0.65]$   &  5.4  & 61  &  -1.64   &   $[-2.13,-1.05]$ & $[-2.52,-0.49]$   & 3.2 & 32 \\
\hline
$\C{9\mu}^{\rm NP}=-3 \C{9e}^{\rm NP}$ &-1.07 & [-1.24,-0.90] & [-1.40,-0.72] & 5.8 & 70 & -1.35  &    $[-1.82,-0.95]$ & $[-2.38,-0.59]$   &4.0 & 72
\end{tabular}
\caption{Most prominent patterns of New Physics in $b\to s\mu\mu$ under the 1D hypothesis. The $p$-values are quoted in \% and Pull$_{\rm SM}$ in units of standard deviation.}
\label{tab:results1D}
\end{table*}

Regarding the theory computation of all observables, we follow Refs.~\cite{Descotes-Genon:2014uoa,Descotes-Genon:2015uva}, which take into account the theoretical updates for the branching ratios of $B\to X_s\gamma$, $B\to X_s\mu\mu$ and $B_s\to \mu\mu$ in Refs.~\cite{Misiak:2015xwa,Huber:2015sra,Bobeth:2013uxa}.
For the $B\to K^\star$ form factors at large recoil we use the calculation in Ref.~\cite{Khodjamirian:2010vf}, which has more conservative uncertainties than the ones in Ref.~\cite{Straub:2015ica}, obtained with a different method. For $B_s\to \phi$ the corresponding calculation is not available, and therefore we use Ref.~\cite{Straub:2015ica}. This leads to smaller hadronic uncertainties quoted for $B_s\to \phi\ell\ell$ and $R_\phi$, but we stress that this is only due to the choice of input. 

We follow the same statistical method as in Ref.~\cite{Descotes-Genon:2015uva}. We perform a frequentist analysis with all known theory and experimental correlations taken into account through the covariance matrix when building the $\chi^2$ function, which is minimised to find best-fit points, pulls, $p$-values and confidence-level intervals. Depending on the dimensionality of the hypothesis, the minimisation is performed either using a simple scan or the Markov-Chain Monte Carlo Metropolis-Hastings algorithm.

\section{Results}

\subsection{Fit results}\label{sec:results}

\begin{table*}[t] 
\begin{tabular}{c||c|c|c||c|c|c}
 & \multicolumn{3}{c||}{All} &  \multicolumn{3}{c}{LFUV}\\
\hline
 2D Hyp.  & Best fit  & Pull$_{\rm SM}$ & p-value & Best fit & Pull$_{\rm SM}$ & p-value\\
\hline\hline
$(\C{9\mu}^{\rm NP},\C{10\mu}^{\rm NP})$ & (-1.01,0.29) & 5.7      &  72 &  (-1.30,0.36)    & 3.7 & 75 \\
$(\C{9\mu}^{\rm NP},\C{7}^{\prime})$  & (-1.13,0.01) &  5.5     &  69 &  (-1.85,-0.04)   & 3.6  &  66\\
$(\C{9\mu}^{\rm NP},\C{9^\prime\mu})$  &  (-1.15,0.41) &  5.6      &  71 & (-1.99,0.93)    &  3.7  &  72\\
$(\C{9\mu}^{\rm NP},\C{10^\prime\mu})$  &   (-1.22,-0.22)     &  5.7  &  72 & (-2.22,-0.41)  & 3.9   & 85 \\ \hline
$(\C{9\mu}^{\rm NP}, \C{9e}^{\rm NP})$ & (-1.00,0.42) &   5.5   &  68 & (-1.36,0.46)  &  3.5  &  65\\ \hline
Hyp. 1  & (-1.16,0.38)   & 5.7      &  73  & (-1.68,0.60)    & 3.8   & 78 \\
Hyp. 2  & (-1.15, 0.01)  &  5.0     &  57  & (-2.16,0.41)    & 3.0    & 37 \\
Hyp. 3  & (-0.67,-0.10)  & 5.0      &  57  &  (0.61,2.48)   & 3.7   &  73 \\
Hyp. 4  & (-0.70,0.28) &  5.0    & 57  &  (-0.74,0.43)   & 3.7 & 72 \\
\end{tabular}
\caption{Most prominent patterns of New Physics in $b\to s\mu\mu$ with high significances. The last four rows corresponds to hypothesis 1: $(\C{9\mu}^{\rm NP}=-\C{9^\prime\mu} , \C{10\mu}^{\rm NP}=\C{10^\prime\mu})$,  2:
 $(\C{9\mu}^{\rm NP}=-\C{9^\prime\mu} , \C{10\mu}^{\rm NP}=-\C{10^\prime\mu})$, 3:
$(\C{9\mu}^{\rm NP}=-\C{10\mu}^{\rm NP} , \C{9^\prime\mu}=\C{10^\prime\mu}$) and 4: $(\C{9\mu}^{\rm NP}=-\C{10\mu}^{\rm NP} , \C{9^\prime\mu}=-\C{10^\prime\mu}$). The ``All'' columns include all available data from LHCb, Belle, ATLAS and CMS, whereas the ``LFUV'' columns are restricted to $R_K$, $R_{K^*}$ and $Q_{4,5}$ (see text for more detail). The $p$-values are quoted in \% and Pull$_{\rm SM}$ in units of standard deviation.}
\label{tab:results2D}\end{table*}

\begin{table*}[t]
\begin{tabular}{c||c|c|c|c|c|c}
 & $\C7^{\rm NP}$ & $\C{9\mu}^{\rm NP}$ & $\C{10\mu}^{\rm NP}$ & $\C{7^\prime}$ & $\C{9^\prime \mu}$ & $\C{10^\prime \mu}$  \\
\hline\hline
Best fit & +0.03 & -1.12 & +0.31 & +0.03 & +0.38 & +0.02 \\ \hline
1 $\sigma$ & $[-0.01,+0.05]$ & $[-1.34,-0.88]$ & $[+0.10,+0.57]$ & $[+0.00,+0.06]$ & $[-0.17,+1.04]$ &$[-0.28,+0.36]$ \\
2 $\sigma$ & $[-0.03,+0.07]$ & $[-1.54,-0.63]$ & $[-0.08,+0.84]$ & $[-0.02,+0.08]$ & $[-0.59,+1.58]$ &$[-0.54,+0.68]$
\end{tabular}
\caption{1 and 2~$\sigma$ confidence intervals for the NP contributions to Wilson coefficients in
the six-dimensional hypothesis allowing for NP in $b\to s\mu\mu$ operators dominant in the SM and their chirally-flipped counterparts, for the fit ``All''. The SM pull is 5.0~$\sigma$.}
\label{tab:Fit6D}
\end{table*}

In Tabs.~\ref{tab:results1D} and \ref{tab:results2D}, we give the fit results for several one- or two-dimensional hypothesis for NP contributions to the various operators, with two different datasets: either we include all available data from muon and electron channels presented in the previous section (column ``All'', 175 measurements), or we include only LFUV observables, i.e., $R_K$ and $R_{K^*}$ from LHCb and $Q_i$ ($i=4,5$) from Belle
(column ``LFUV'', 17 measurements). In both cases, we include also the $b\to s\gamma$ observables, as well as ${\cal B}(B\to X_s\mu\mu)$ and ${\cal B}(B_s\to\mu\mu)$. The SM point yields a $\chi^2$ corresponding to a $p$-value of 11.3\% for the fit ``All'' and 4.4\% for the fit ``LFUV''.

We start by discussing NP hypotheses for the fit ``All''. The measurement of $R_{K^*}$ increases further the significance of already prominent hypotheses in previous studies,  namely, the first three hypotheses ($\C{9\mu}^{\rm NP}$, $\C{9\mu}^{\rm NP}=-\C{10\mu}^{\rm NP}$ and $\C{9\mu}^{\rm NP}=-\C{9^\prime\mu}$) already identified in Refs.~\cite{Descotes-Genon:2013wba,Descotes-Genon:2015uva}. The SM pull exceeds 5~$\sigma$ in each case: the hypotheses can hardly be distinguished on this criterion, and as discussed in Ref.~\cite{Capdevila:2016ivx}, the $Q_i$ observables will be very powerful tools to lift this quasi-degeneracy.

Besides providing the results for one- and two-dimensional hypotheses with SM pulls above $5\,\sigma$, we discuss four illustrative examples of NP hypotheses with specific chiral structures, leading to correlated shifts in Wilson coefficients. These hypotheses are:\\[3mm]
1. $(\C{9\mu}^{\rm NP}=-\C{9^\prime\mu} , \C{10\mu}^{\rm NP}=\C{10^\prime\mu})$,\\[3mm]
2. $(\C{9\mu}^{\rm NP}=-\C{9^\prime\mu} , \C{10\mu}^{\rm NP}=-\C{10^\prime\mu})$,\\[3mm]
3. $(\C{9\mu}^{\rm NP}=-\C{10\mu}^{\rm NP} , \C{9^\prime\mu}=\C{10^\prime\mu}$),\\[3mm]
4. $(\C{9\mu}^{\rm NP}=-\C{10\mu}^{\rm NP} , \C{9^\prime\mu}=-\C{10^\prime\mu}$).\\[3mm]

Hypothesis 1 has the highest SM pull, in agreement with our previous global analysis \cite{Descotes-Genon:2015uva}. Taking $\C{10\mu}^{\rm NP}=-\C{10^\prime\mu}$ (i.e., Hypothesis 2) reduces the significance from $5.7\,\sigma$ to $5.0\,\sigma$, similarly to Hypotheses 3 and 4 taking $\C{9\mu}^{\rm NP}=-\C{10\mu}^{\rm NP}$ (irrespectively of the relative sign taken to constrain $C_{9^\prime\mu}=\pm C_{10^\prime\mu}$). 
From a model-independent point of view, Hypothesis 1 is particularly interesting to yield a low value for $R_{K^*}$ (especially if a contribution $\C7^{\rm NP}>0$ is allowed).
Let us add that  a scenario with only $\C{9\mu}^{\rm NP}=-\C{9^\prime\mu}$ would predict $R_K=1$ and $R_{K^*} < 1$~\cite{Hiller:2014ula,Hiller:2014yaa,Descotes-Genon:2015uva}.  One could however  obtain $R_K<1$ by adding a  positive contribution to $\C{10\mu}$ and/or $\C{10^\prime\mu}$ (see Tab.~9 in Ref.~\cite{Descotes-Genon:2015uva}). 
 
Up to now, we have discussed scenarios where NP contributions occur only in $b\to s\mu\mu$ transitions. It is also interesting to consider scenarios with NP in both muon and electron channels, in particular $(\C{9\mu}^{\rm NP},\C{9e}^{\rm NP})$, with a SM pull of $5.5\,\sigma$, and a  $p$-value of 68\%. While $\C{9\mu}^{\rm NP}\sim -1$ is preferred over the SM with a significance around $5\,\sigma$, $\C{9e}$ is compatible with the SM already at $1\,\sigma$, in agreement with the LFUV data included in the fit. One can assess more precisely the need for LFUV in the framework where NP is allowed in both $(\C{9e}^{\rm NP}$ and $\C{9\mu}^{\rm NP}$) through the pull of the hypothesis $(\C{9e}^{\rm NP}=\C{9\mu}^{\rm NP})$ which reaches $3.3\,\sigma$. Considering the results for the $(\C{9e}^{\rm NP},\C{9\mu}^{\rm NP})$ hypothesis, one can notice that a very good fit is also obtained for the one-dimensional hypothesis $\C{9\mu}^{\rm NP}=-3 \C{9e}^{\rm NP}$ favoured in some models discussed in the next section.

\begin{figure*}
\begin{center}
\includegraphics[width=0.32\textwidth]{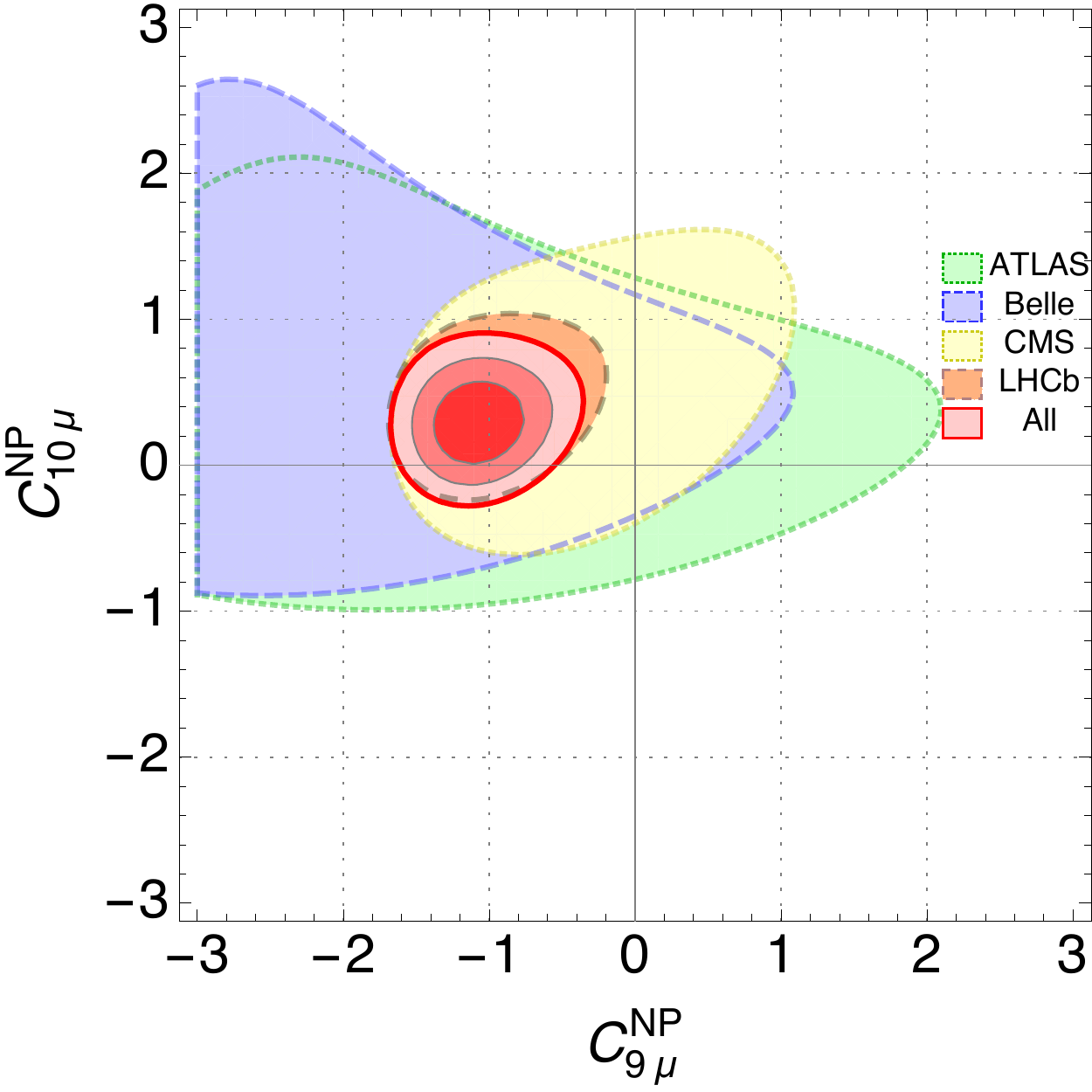}\hspace{2mm}
\includegraphics[width=0.32\textwidth]{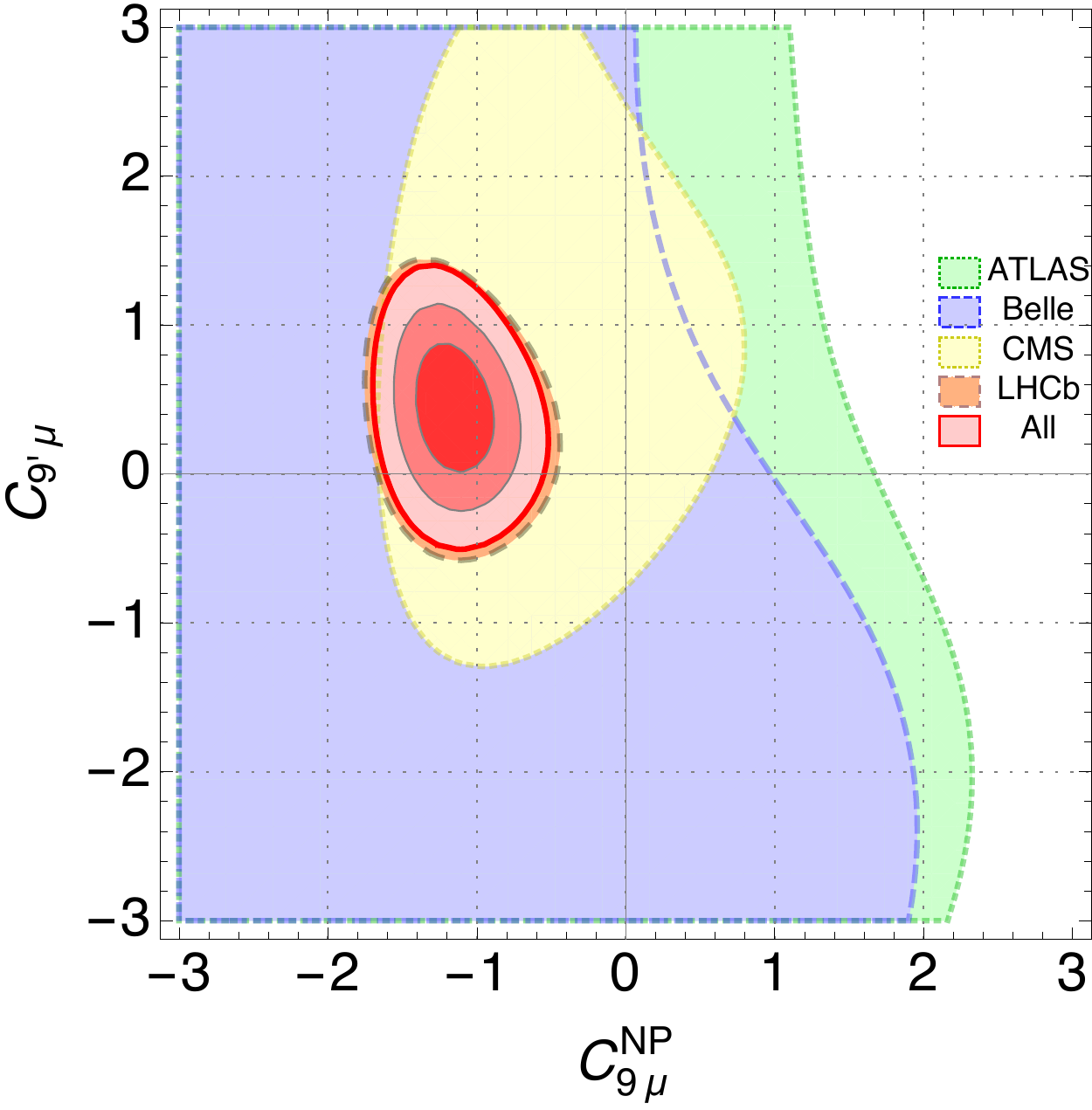}\hspace{2mm}
\includegraphics[width=0.32\textwidth]{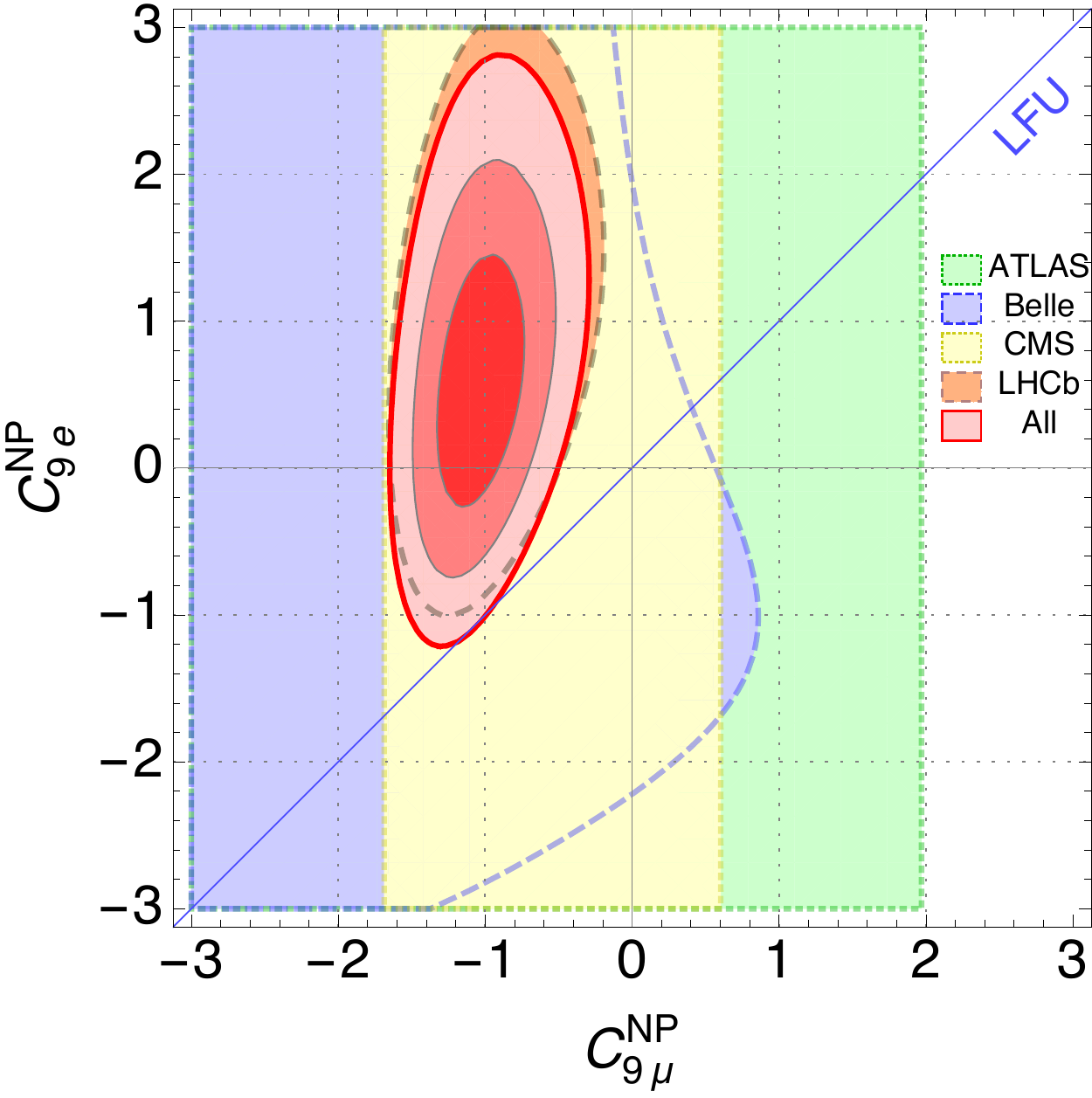}
\end{center}
\caption{From left to right: Allowed regions in the $(\C{9\mu}^{\rm NP},\C{10\mu}^{\rm NP})$, $(\C{9\mu}^{\rm NP},\C{9^\prime\mu})$ and $(\C{9\mu}^{\rm NP},\C{9e}^{\rm NP})$ planes for the corresponding two-dimensional hypotheses, using all available data (fit ``All''). We also show the 3 $\sigma$ regions for the data subsets corresponding to specific experiments. Constraints from $b\to s\gamma$ observables, ${\cal B}(B\to X_s\mu\mu)$ and ${\cal B}(B_s\to \mu\mu)$ are included in each case (see text).}         
\label{fig:FitResultAll}
\end{figure*}

\begin{figure*}
\begin{center}
\includegraphics[width=0.32\textwidth]{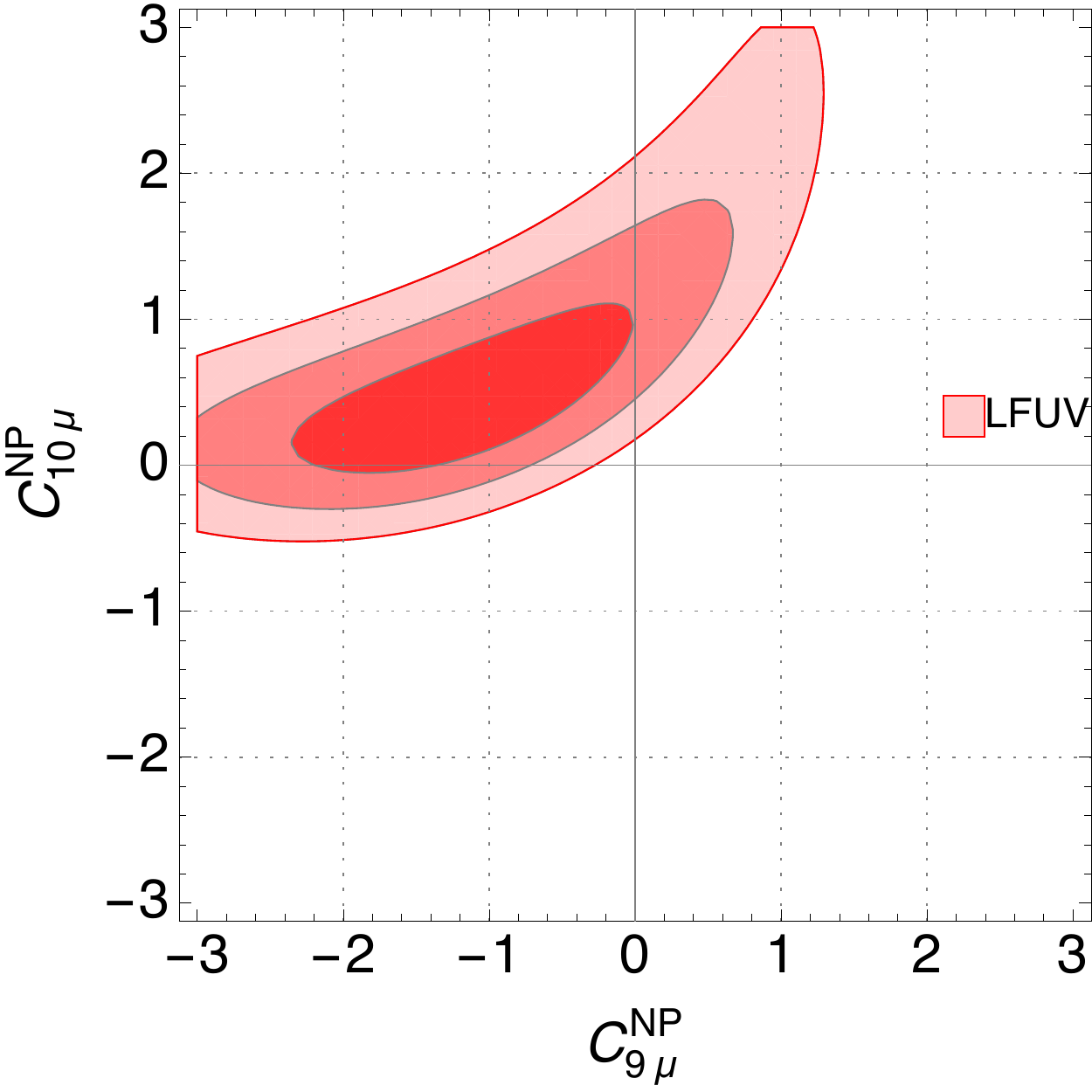}\hspace{2mm}
\includegraphics[width=0.32\textwidth]{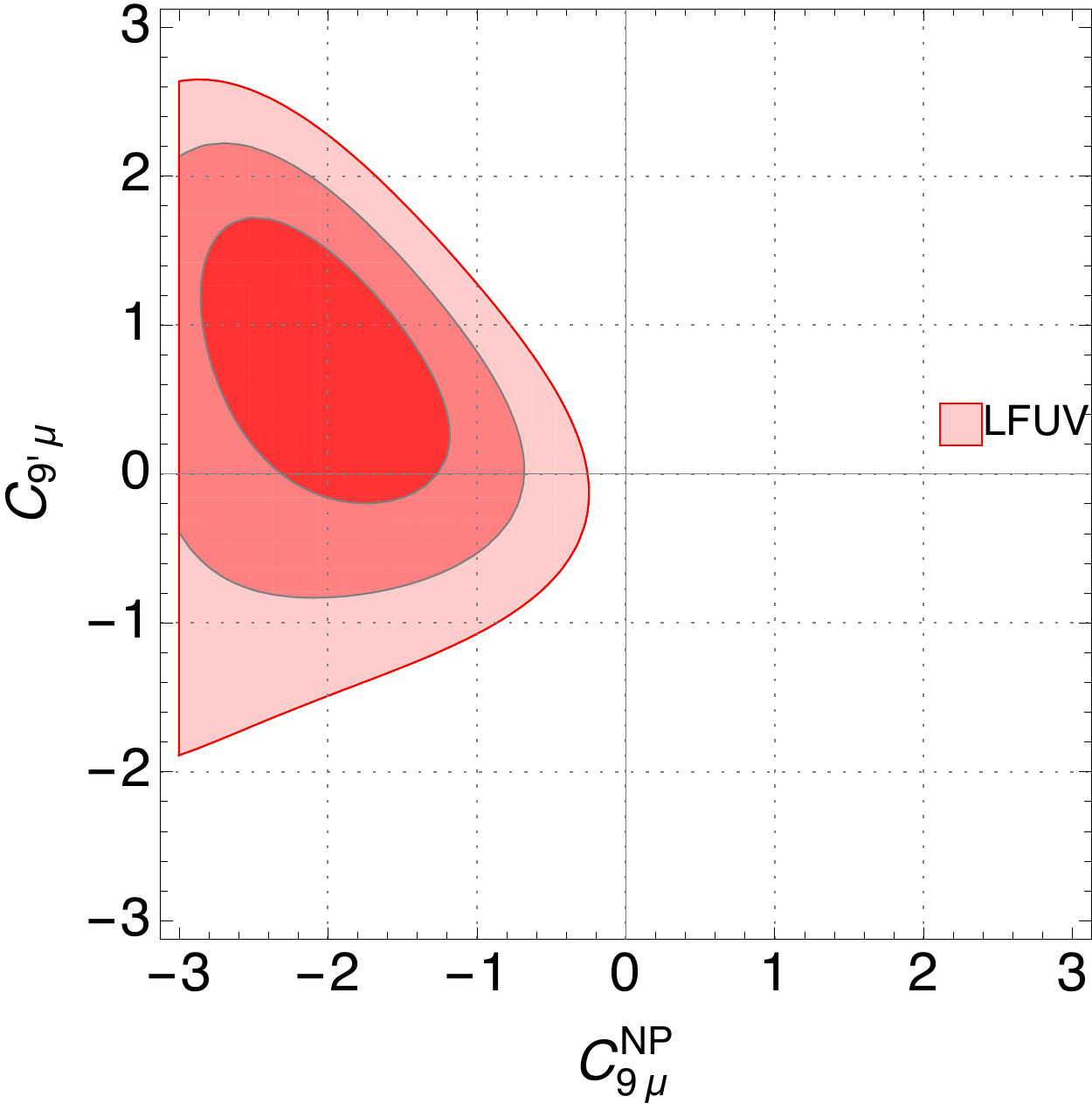}\hspace{2mm}
\includegraphics[width=0.32\textwidth]{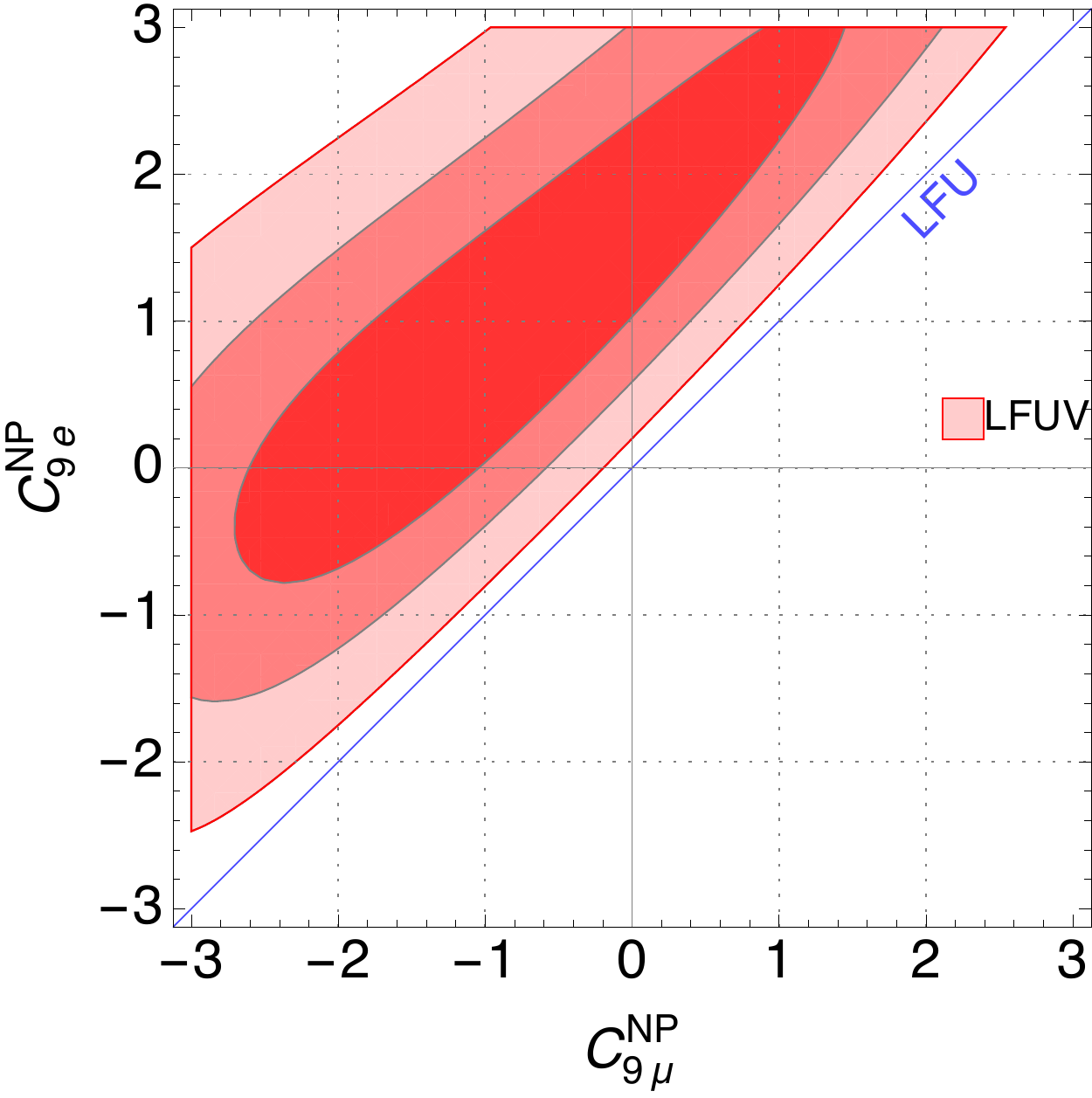}
\end{center}
\caption{From left to right: Allowed regions in the $(\C{9\mu}^{\rm NP},\C{10\mu}^{\rm NP})$, $(\C{9\mu}^{\rm NP},\C{9^\prime\mu})$ and $(\C{9\mu}^{\rm NP},\C{9e}^{\rm NP})$ planes for the corresponding two-dimensional hypotheses, using only LFUV observables (fit ``LFUV''). Constraints from $b\to s\gamma$ observables, ${\cal B}(B\to X_s\mu\mu)$ and ${\cal B}(B_s\to \mu\mu)$ are included in each case (see text).}         
\label{fig:FitResultLFUV}
\end{figure*}

In Fig.~\ref{fig:FitResultAll} we show the corresponding constraints for the fit ``All'' under the three hypotheses $(\C{9\mu}^{\rm NP},\C{10\mu}^{\rm NP})$, $(\C{9\mu}^{\rm NP},\C{9\mu^\prime})$ and $(\C{9\mu}^{\rm NP},\C{9e}^{\rm NP})$, as well as the $3\,\sigma$ regions according to the results from individual experiments (for each region, we add the constraints from $b\to s\gamma$ observables, ${\cal B}(B\to X_s\mu\mu)$ and the world average for ${\cal B}(B_s\to \mu\mu)$~\cite{Amhis:2016xyh}). As expected, the LHCb results drive most of the effect, with a clear exclusion of the origin, i.e., the SM point. 

We can now move to the fit ``LFUV'' in Fig.~\ref{fig:FitResultLFUV}, where we consider the same hypotheses favoured by global analyses. It is interesting to notice that this restricted subset of observables excludes the SM point with a high significance, and it favours regions similar to the fit ``All'' dominated by different $b\to s\mu\mu$-related observables ($B\to K^*\mu\mu$ optimised angular observables as well as low- and large-recoil branching ratios for $B\to K\mu\mu$, $B\to K^*\mu\mu$ and $B_s\to\phi\mu\mu$). This  is also shown in Tabs.~\ref{tab:results1D} and \ref{tab:results2D}, where the scenarios with the highest pulls are confirmed with significances between 3 and 4 $\sigma$, but get harder to distinguish on the basis of their significance. Scenarios like $\C{9\mu}^{\rm NP}=-\C{9^\prime\mu}$ that would fail to explain $R_K$ are not disfavoured due to their good compatibility with $R_{K^{*}}$ data. Interestingly, the inclusion of the $R_{K^*}$ measurement now disfavours solutions with right-handed currents only, as proposed in Ref.~\cite{Becirevic:2015asa,Becirevic:2016yqi}. Such a scenario was valid considering only $R_{K}$ (excluding the other $b\to s\mu^+\mu^-$ data), but is now disfavoured by the measurement of $R_{K^*}$. This was solved later on in \cite{Becirevic:2017jtw},  by modifying the model via a scalar leptoquark with hypercharge $Y=7/6$.  

Finally, we have performed a six-dimensional fit allowing for NP contributions in $\C{7('),9(')\mu,10(')\mu}$. The SM pull has shifted from 3.6$\sigma$ in the fit of Ref.~\cite{Descotes-Genon:2015uva}
to 5.0~$\sigma$ if one considers the fit ``All'' described above.
The 1 and 2~$\sigma$ CL intervals are given in Tab.~\ref{tab:Fit6D}, with the pattern:
\begin{equation}
\C7^{\rm NP}\gtrsim 0,   \, \C{9\mu}^{\rm NP}<0  ,\, \C{10\mu}^{\rm NP} >0 ,\, \C{7^\prime} \gtrsim 0 ,\, \C{9^{\prime} \mu}>0  ,\, \C{10^{\prime}\mu} \gtrsim 0
\end{equation}
where $\C{9\mu}$ is compatible with the SM beyond 3~$\sigma$, $\C{10\mu}$, $\C{7^\prime}$  at 2~$\sigma$ and all the other coefficients at 1~$\sigma$.

\subsection{$R_{K}$ and $R_{K^*}$: A closer look}
\label{sec:RKRD}

Theoretical predictions in the SM for $R_K$ and $R_{K^*}$ are very accurate: hadronic uncertainties cancel to a large extent and electromagnetic corrections have been estimated to be small and under control~\cite{Bordone:2016gaq}.
This is true as long as there are no significant LFUV effects. If there are, interference effects between LFUV and LFU conserving contributions spoil the cancellation of hadronic uncertainties.
These effects might come from NP or from lepton-mass effects in the SM. The latter are only important at very low $q^2$,
wherever $m_\ell^2/q^2$ is not small compared to 1 (say, below $q^2 \sim 1$GeV$^2$), and affect in particular the first measured bin
in $R_{K^*}$. In this bin one thus expects larger theoretical uncertainties than in the region above 1~GeV$^2$, as well
as at any value of $q^2$ in the presence of LFUV new physics~\cite{Capdevila:2017ert,Capdevila:2016ivx}.
This enhancement of the uncertainty is less important in the optimized LFUV observables $Q_i$~\cite{Capdevila:2016ivx}.
An exception to this enhancement occurs under  the hypothesis $\C{9\mu}^{\rm NP}=-\C{10\mu}^{\rm NP}$: above 1 GeV$^2$, the contribution of right-handed amplitudes to $R_{K^*}$ cancel to a large extent, reducing the theoretical uncertainty substantially.

Large-recoil expressions for the transversity amplitudes  can be used  to provide approximate expressions for $R_{K^*}$ in the first two bins in terms of Wilson coefficients, leading to further cross-checks of our predictions.  Let us stress that the following approximate expressions are given for illustrative purposes, and that complete expressions have been used for all the numerical evaluations in this article (see also Refs.~\cite{Capdevila:2016ivx} and \cite{Capdevila:2017ert} for exact predictions). We consider the large-recoil limit and we work under the hypothesis that New Physics enters in muon modes and is suppressed for electrons \cite{Ghosh:2014awa, Descotes-Genon:2015uva}. In the first bin one finds:
$$R_{K^*}^{[0.045,1.1]}\simeq \left(12.8+g_{(1)}^\mu+g_{(2)}^\mu \right)/\left(13.4+g_{(1)}^e + g_{(2)}^e \right)$$
where $g_{(i)}^\ell$ stands for the linear ($i=1$) and quadratic ($i=2$) term for $\ell=e,\mu$ and are given by:
\begin{eqnarray} g_{(1)}^\ell=&&-1.1\left[ \C{10\ell}^{\rm NP} - \C{9\ell}^{\rm NP} /2 + \C{9^\prime \ell} -\C{10^\prime \ell} \right] 
\nonumber \\ &&-61.9\, \C7^{\rm NP}-1.7\, \C7^\prime\ ,
\end{eqnarray} 
and
\begin{eqnarray} g_{(2)}^\ell&\!\!=\!\!& 
 -0.7\,  \C7^{\rm NP} \C7^\prime +123.1 \left[ (\C7^{\rm NP})^ 2+(\C{7^\prime})^2\right]  \nonumber \\ 
&&\hspace{-3.5mm} +2.2 \left[ \C7^{\rm NP} \C{9\ell}^{\rm NP} + \C{7^\prime} \C{9^\prime \ell} \right] \\
&&\hspace{-3.5mm} + 0.1 \left[ (\C{9\ell}^{\rm NP})^2+ (\C{10\ell}^{\rm NP})^2 + (\C{9^\prime\ell})^2+ (\C{10^\prime\ell})^2 \right] \nonumber \\
&&\hspace{-3.5mm} -0.4 \left[ \C7^{\rm NP} \C{9^\prime \ell} + \C7^\prime \C{9\ell}^{\rm NP}+ \frac{1}{2}(\C{9\ell}^{\rm NP} \C{9^\prime \ell} + \C{10\ell}^{\rm NP} \C{10^\prime \ell}) \right] \nonumber
\end{eqnarray}
showing that a negative (positive) contribution to $\C{9\mu}^{\rm NP}$ and $\C{10^\prime \mu}$ ($\C{10\mu}^{\rm NP}$ and $\C{9^\prime\mu}$) enhances the deviation from SM. The (universal) radiative coefficients $\C{7}$ and $\C{7^\prime}$ play also a (subleading) role in mixed terms combining them with the semileptonic NP coefficients in this bin. 

In the second bin, the expression gets simplified due to the very limited impact on $R_{K^*}$ of the radiative coefficients $\C{7}$ and $\C{7^\prime}$ (not shown here):
\begin{equation}
R_{K^*}^{[1.1,6]}\simeq \left(29.2+\tilde g_{(1)}^\mu+ \tilde g_{(2)}^\mu \right)/\left(29.3+ \tilde g_{(1)}^e + \tilde g_{(2)}^e \right)
\end{equation}
with
\begin{equation} \tilde g_{(1)}^\ell
   = -8.1\, \C{10\ell}^{\rm NP} +5.3\, \C{10^\prime \ell} + 5.6\, \C{9\ell}^{\rm NP} -5.0\, \C{9^\prime \ell}
\end{equation}
and
\begin{eqnarray}
 \tilde g_{(2)}^\ell=&& + 0.9 \left[(\C{10\ell}^{\rm NP})2 + (\C{10^\prime\ell})^2 + (\C{9\ell}^{\rm NP})^2 
  + (\C{9^\prime \mu})^2 \right] \nonumber \\&&-1.2 \left[ \C{9\ell}^{\rm NP} \C{9^\prime \ell} + \C{10\ell}^{\rm NP} \C{10^\prime \ell} \right] 
 \end{eqnarray}
In the presence of NP,  the same mechanisms as in the first bin operate here, but with a stronger impact.

 A last comment is in order concerning the relatively low value of $R_{K^*}$ in the first bin. It is difficult to accommodate a very low value  of $R_{K^*}$ in this first bin through NP contributions to semileptonic $\C{9\mu}, \C{10\mu}$ coefficients (in agreement with the fit), since the branching ratio in this region is dominated by LFU operator ${\cal{O}}_{7}$ (the photon pole). A low value can be obtained if a positive contribution $\C7^{\rm NP}=O(0.1)$ is added together with a small positive (negative) contribution to $\C{9^\prime\mu}$ ($\C{10^\prime\mu}$), but such a large contribution is however not favored by $b\to s\gamma$ observables. Moreover, the second bin will be even lower than the first one. It seems thus likely that the very low value of the first bin for $R_{K^*}$ is partly due to a downward statistical fluctuation. We will not dwell on this issue further
and we let the fit resolve whether this leads to significant tensions.

\subsection{Implications for models}\label{sec:models}

Our updated model-independent fit to available $b\to s\ell\ell$ and $b\to s\gamma$ data strongly favours LFUV scenarios with NP affecting mainly $b\to s\mu\mu$ transitions, with a preference for the three hypotheses $\C{9\mu}^{\rm NP}$, $\C{9\mu}^{\rm NP}=-\C{10\mu}^{\rm NP}$ and $\C{9\mu}^{\rm NP}=-\C{9^\prime\mu}$. This has important implications for some popular ultraviolet-complete models which we briefly discuss.\\[-2mm]

\noindent $\blacktriangleright$ {\bf LFUV}: Given that leptoquarks (LQs) should posses very small couplings to electrons in order to avoid dangerous effects in $\mu\to e\gamma$, they naturally violate LFU. While $Z^\prime$ models can easily accommodate LFUV data~\cite{Falkowski:2015zwa}, LFU variants like the ones in Refs.~\cite{Gauld:2013qba,Buras:2013dea} are now disfavoured. The same is true if one aims at explaining $P_5'$ via NP in four-quark operators leading to a NP ($q^2$-dependent) contribution from charm loops~\cite{Jager:2017gal}. As already discussed, models with right-handed currents such as Refs.~\cite{Becirevic:2016yqi,Cox:2016epl} are also strongly disfavoured, even though they can account for $R_K$, since they would result in $R_{K^*}>1$.\\[-2mm]

\noindent $\blacktriangleright$ {\begin{boldmath} $\C{9\mu}^{\rm NP}$: \end{boldmath}} $Z^\prime$ models with fundamental (gauge) couplings to leptons preferably yield $\C{9\mu}^{\rm NP}$-like solutions in order to avoid gauge anomalies. In this context, $L_\mu-L_\tau$ models~\cite{Altmannshofer:2014cfa,Crivellin:2015mga,Crivellin:2015lwa,Crivellin:2016ejn} are popular since they do not generate effects in electron channels. The new fit including $R_{K^*}$ is also very favourable to models predicting $\C{9\mu}^{\rm NP}=-3\C{9e}^{\rm NP}$~\cite{Bhatia:2017tgo}. Interestingly, such a symmetry pattern is in good agreement with the structure of the PMNS matrix. Concerning LQs, a $\C{9\mu}^{\rm NP}$-like solution can only be generated by adding two scalar (an $SU(2)_L$ triplet and an $SU(2)_L$ doublet with $Y=7/6$) or two vector representations (an $SU(2)_L$ singlet with $Y=2/3$ and an $SU(2)_L$ doublet with $Y=5/6$).\\[-2mm] 

\noindent $\blacktriangleright$ {\begin{boldmath} $\C{9\mu}^{\rm NP}=-\C{10\mu}^{\rm NP}$: \end{boldmath}} This pattern can be achieved in $Z^\prime$ models with loop-induced couplings~\cite{Belanger:2015nma} or in $Z^\prime$ models with heavy vector-like fermions~\cite{Boucenna:2016wpr,Boucenna:2016qad} which posses also LFUV. Concerning LQs, here a single representation (the scalar $SU(2)_L$ triplet or the vector $SU(2)_L$ singlet with $Y=2/3$) can generate a $\C{9\mu}=-\C{10\mu}$ like solution~\cite{Gripaios:2014tna,Fajfer:2015ycq,Varzielas:2015iva,Alonso:2015sja,Calibbi:2015kma,Barbieri:2015yvd,Sahoo:2016pet} and this pattern can also be obtained in models with loop contributions from three heavy new scalars and fermions~\cite{Gripaios:2015gra,Arnan:2016cpy,Mahmoudi:2014mja}. Composite Higgs models are also able to achieve this pattern of deviations~\cite{Niehoff:2015bfa}.\\[-2mm] 

\noindent $\blacktriangleright$ {\begin{boldmath} $\C{9\mu}^{\rm NP}=-\C{9^\prime\mu}$: \end{boldmath}} This pattern could be generated in $Z^\prime$ models with vector-like fermions. For the $L_\mu-L_\tau$ model~\cite{Altmannshofer:2014cfa} this would be naturally the case if vector-like fermions and the generalized Yukawa couplings respect a left-right symmetry. One could also obtain this pattern by adding a third Higgs doublet to the model of Ref.~\cite{Crivellin:2015lwa} with opposite $U(1)$ charge. Generating $\C{9\mu}^{\rm NP}=-\C{9^\prime\mu}^{\rm NP}$ in LQ models requires one to add four scalar representations or three vector ones.\\[-2mm] 

Concerning the constrained 2D hypotheses in the lower part of Tab.~\ref{tab:results2D}, only two of them (2 and 4) can be explained within a $Z^\prime$ model, while hypotheses 1 and 3 violate the relationship $\C{9\mu}^{\rm NP} \times \C{10^\prime\mu}=\C{10\mu}^{\rm NP} \times \C{9^\prime\mu}$ \cite{Descotes-Genon:2015uva} that minimal $Z^\prime$ models should obey. One would have to turn to other models (like LQs with a sufficient number of representations) to explain the hypothesis with the highest pull (Hyp. 1).

We close the section by correlating the violation of lepton flavour universality observed in $b\to s\ell\ell$ with the  measurements of $R_D$ and $R_{D^*}$ that also point towards LFUV with a combined significance of $3.9\,\sigma$~\cite{Amhis:2016xyh}. Such a correlation between $b\to s$ and $b\to c$ transitions, however, requires further hypotheses.  A solution of the $R_{D^{(*)}}$ anomaly can naturally be achieved with a NP contribution to the SM operator $\bar c\gamma^\mu P_L b \bar \tau\gamma_\mu P_L \nu$ as it complies with the $B_c$ lifetime~\cite{Alonso:2016oyd} and $q^2$ distributions~\cite{Freytsis:2015qca,Celis:2016azn,Ivanov:2017mrj}. 
 Assuming $SU(2)$ invariance, the effect in  $R_{D^{(*)}}$ is correlated to $b\to s\ell^+\ell^-$ and/or to $b\to s\nu\bar\nu$, following the pattern $\C{9\mu}=-\C{10\mu}$. Following model-independent arguments, $b\to s\tau^+\tau^-$ must then be significantly enhanced. Indeed, since $b\to c\ell\nu$ processes are mediated already at tree level in the SM, a rather large NP contribution is required and in principle large contributions to $b\to s\nu\bar\nu$ processes appear, due to $SU(2)$ invariance. These bounds from $B\to K^{(*)}\nu\bar\nu$ can be avoided if the coupling structure is mainly aligned to the third generation, but this disagrees with direct LHC searches~\cite{Faroughy:2016osc} and electroweak precision observables~\cite{Feruglio:2016gvd}. However, there is no effect in $b\to s\nu\bar\nu$ processes in the case of a contribution $\C{1}^{\rm NP}=\C3^{\rm NP}$ to gauge-invariant operators~\cite{Grzadkowski:2010es,Celis:2017doq}, which can be achieved with the vector LQ $SU(2)$ singlet~\cite{Alonso:2015sja,Calibbi:2015kma} or with a combination of two scalar LQs~\cite{Crivellin:2017zlb}. 
 In both cases large effects in $b\to s\tau^+\tau^-$ (of the order of $10^{-3}$ for $B_s\to\tau^+\tau^-$) are predicted~\cite{Crivellin:2017zlb, GarciaGarcia:2016nvr}. 
 
 Assuming that the coupling to the second generation is sizeable in order to avoid the bounds from direct LHC searches and electroweak precision observables one finds 
\begin{equation}
\C{9(10)\tau} \approx \C{9(10)}^{SM} -(+) 2\frac{\pi}{\alpha}\frac{{V_{cb}}}{{V_{ts}^*}}\left( {\sqrt {\frac{R_{D^{(*)}}}{R_{D^{(*)}}^{\rm SM}}}  - 1} \right)\,.
\end{equation}
Furthermore, in LQ models one expects sizeable branching ratios for $b\to s\tau\mu$ processes, reaching $10^{-5}$~\cite{Crivellin:2017zlb}.

\section{A data-driven consistency test of hadronic uncertainties}
\label{sec:hadtest}

\begin{figure}
\includegraphics[width=8.5cm]{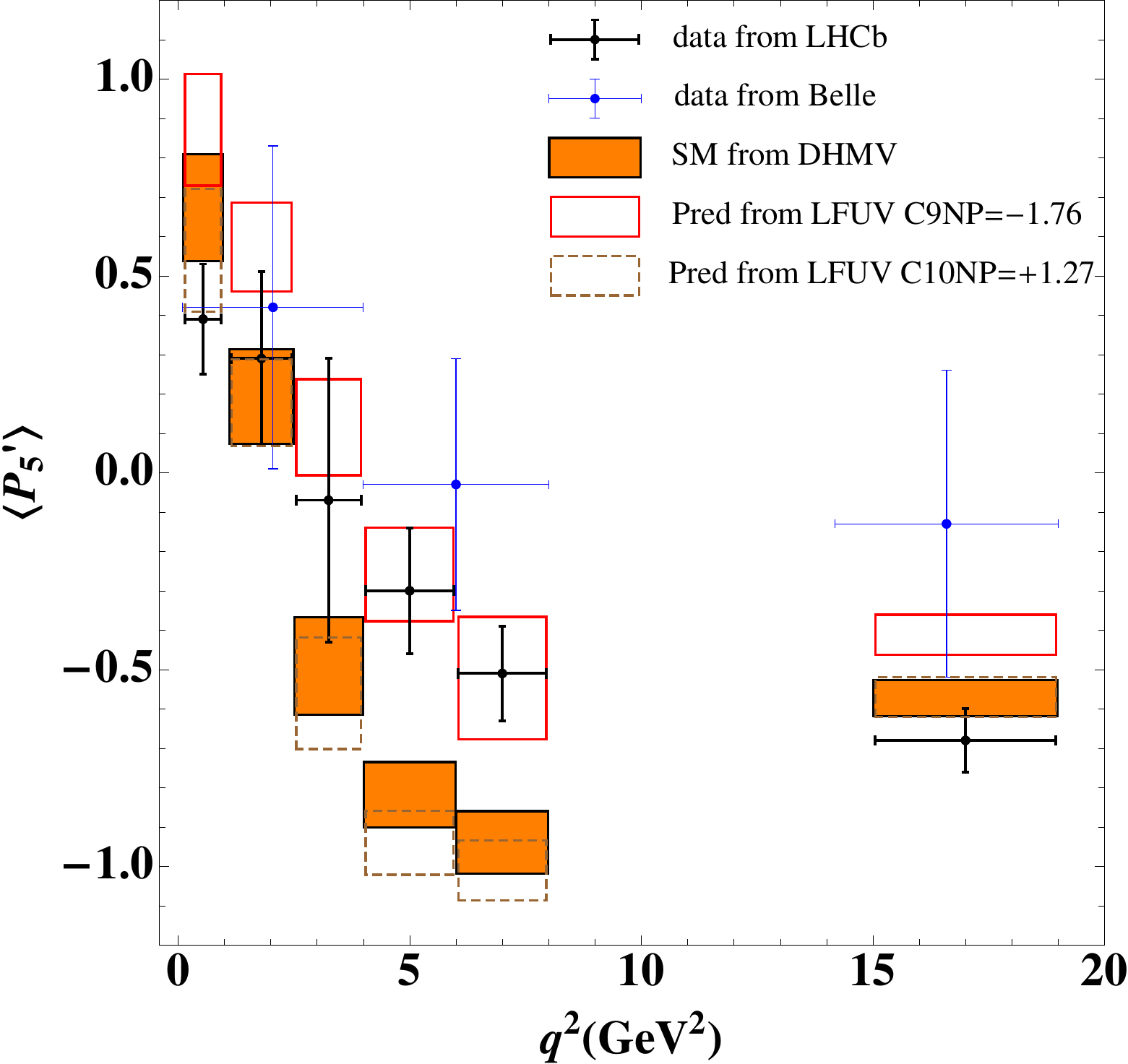}
\caption{Predicted value for $P_5^\prime$   using as input the data from LFUV observables in scenario with NP in $C_{9\mu}=-1.76$ (in red) from present paper or $C_{10\mu}=+1.27$ (in brown). We also give our SM prediction (orange filled boxes) and data from LHCb (black crosses) and Belle (blue crosses).}
\label{fig:plotP5p_pred}
\end{figure} 

\begin{figure}[b]
\mbox{\hspace{-5mm}\includegraphics[height=7.5cm,width=10.1cm]{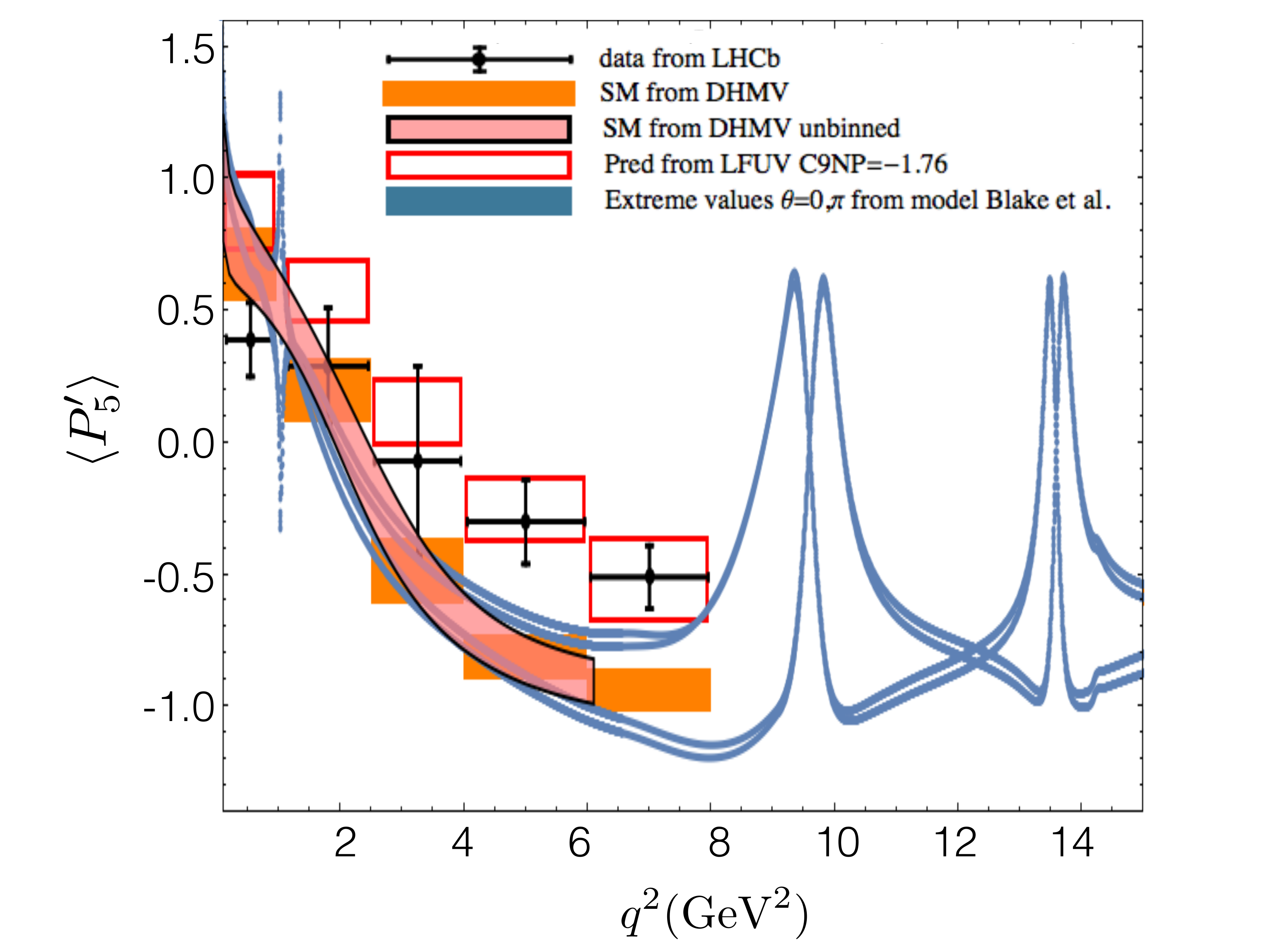}}
\caption{Comparison between the error estimate in $P_5^\prime$ using the empirical model presented in Ref.~\cite{Blake:2017fyh} and our own estimate of $P_5^\prime$ error band in Ref.\cite{Descotes-Genon:2015uva}.
Black crosses represent LHCb data, our SM predictions for $P_5^\prime$ in bin form correspond to the filled orange rectangles, pink band corresponds to our SM prediction for $P_5^\prime$ as a $q^2$ function and blue bands correspond to the extreme values for the strong phase $\theta=0,\pi$ of Ref.~\cite{Blake:2017fyh} within their empirical model. Red rectangles corresponding to the NP in $C_{9\mu}=-1.76$ are also included for completeness.}

\label{fig:comparison}
\end{figure}

The two types of observables included in the fits (LFUV ratios and exclusive $b\to s\ell\ell$ observables) are at a different
level of theoretical control with regards to hadronic uncertainties.
In this sense, there has been an ongoing controversy about the possibility that underestimated hadronic uncertainties may be ultimately responsible for the observed anomalies in $B \to K^*\mu^+\mu^-$ channel, invoking either power corrections to form factors~\cite{Jager:2012uw,Jager:2014rwa} or charm-loop contributions~\cite{Ciuchini:2015qxb,Ciuchini:2016weo}. Even if these arguments have been addressed in detail in Refs.~\cite{Descotes-Genon:2014uoa,Capdevila:2017ert,Aaij:2016cbx,Bobeth:2017vxj}, additional data can help 
to discard them, checking the consistency of the NP deviations and the robustness of our treatment of hadronic uncertainties.

One can consider only the subset of LFUV observables (free from any significant hadronic uncertainties)  to determine the NP contribution to the Wilson coefficient $C_9$. The result shown in Tab.~\ref{tab:results1D} yields a best fit value of $C_{9\mu}^{\rm NP}=-1.76$.  Using this value, one can predict the deviation from SM for $P_5^\prime$, using only the data from LFUV observables. The result shown in Fig.~\ref{fig:plotP5p_pred} shows a remarkable agreement between the predicted value using LFUV observables only and the measurement by LHCb in the bins [4,6] and [6,8]. 

This simple but powerful test supports both that the patterns of deviations are related between LFUV observables and $P_5^\prime$, and that the methodology used to treat hadronic uncertainties in $P_5^\prime$ is appropriate. Conversely, it gives little room for alternative explanations based on hadronic uncertainties to  explain the deviations observed in $P_5^\prime$, given that hadronic effects making the measurement of $P_5^\prime$ SM-like will introduce a tension with
the measurements of LFUV observables.

The same exercise can be done with other scenarios of New Physics, such as the case \mbox{$C_{10\mu}^{\rm NP}=1.27$},
which would also fit well the $R_{K^{(*)}}$ data.
In this case one can see that the prediction for $P_5^\prime$ goes below the SM prediction, increasing the tension further.

Recently, a group of LHCb experimentalists proposed an empirical model describing the long-distance contribution from charm loops,  through the overlap of $J^{PC}=1^{--}$ resonances and fitting their parameters to LHCb data~\cite{Blake:2017fyh}. We compare this model with our own estimate
of long-distance charm contribution to $P_5^\prime$ in Fig.~\ref{fig:comparison}, showing a very appealing agreement both in central values and uncertainties. These estimates are also in agreement with the results in Ref.~\cite{Bobeth:2017vxj}.

\section{Future opportunities for LFUV}
\label{sec:oppLFUV}

\begin{figure}
\includegraphics[width=8.5cm]{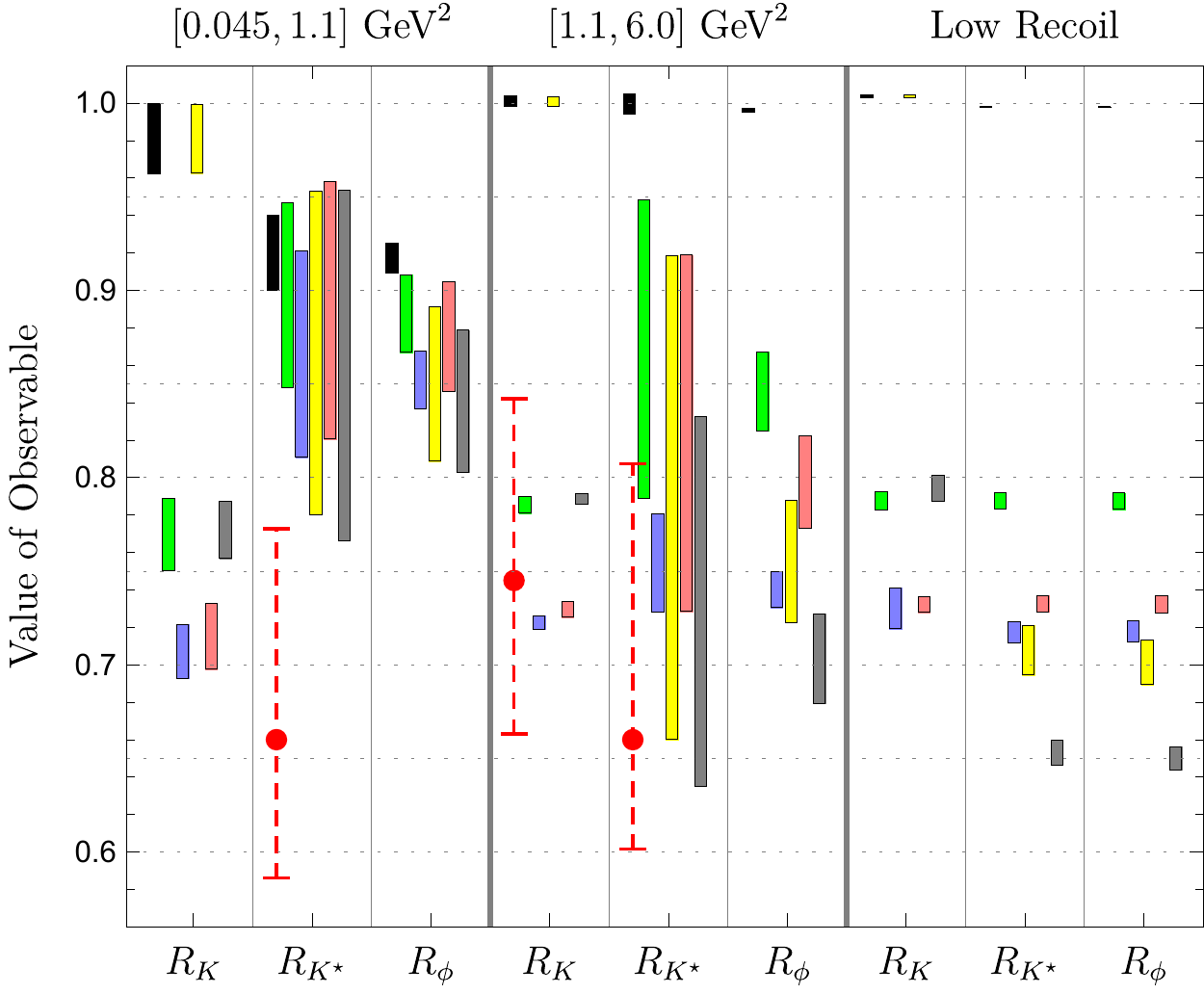}
\caption{Predictions and experimental measurements for $R_K$, $R_{K^*}$ and $R_\phi$ with the same conventions as Fig.~\ref{fig:plotQi}. In the central
box, the predictions for $R_K$ are given for the bin [1,6]~GeV$^2$, whereas $R_{K^*}$ and $R_\phi$ are given in [1.1,6] GeV$^2$. The low-recoil bin corresponds to [15,22]~GeV$^2$, [15,19]~GeV$^2$ or [15,18.8]~GeV$^2$ for $R_K$, $R_{K^\star}$ and $R_\phi$ respectively. The smaller uncertainties in $R_\phi$ (compared to $R_{K^\star}$) is due to the choice of form factors in each case, see Sec.~\ref{sec:framework}. \label{fig:plotR}}
\end{figure}

The best NP scenarios obtained from the global fits have a similar goodness of fit and describe the anomalies with an equivalent success. New measurements will determine eventually which scenario is singled out.
In this respect, a few of the optimised
observables measuring LFUV proposed in Ref.~\cite{Capdevila:2016ivx} are particularly
promising, with pioneering measurements from the Belle experiment  for $Q_{4,5}$~\cite{Wehle:2016yoi}.

In order to illustrate the future potential for establishing which one (if any) of the
various NP scenarios is preferred, we consider not only $R_{K,K^\star,\phi}$ but also the observables  $\hat Q_{1,2,4,5}$
and $B_{5,6s}$ in the same $q^2$ bins as the $R_{K^\star}$ LHCb measurements:
$[0.045,1.1]$, $[1.1,6.0]$ and $[15,19]$~GeV$^2$, and calculate the predictions
within the SM as well as within five promising scenarios considered in the main article:\\[-2mm]

\noindent $\blacktriangleright$ Scenario~1: $C_{9\mu}^{\rm NP} =-1.1$,\\[-2mm]

\noindent $\blacktriangleright$ Scenario~2: $C_{9\mu}^{\rm NP} = -C_{10\mu}^{\rm NP} =-0.62$,\\[-2mm]

\noindent $\blacktriangleright$ Scenario~3: $C_{9\mu}^{\rm NP} = -C'_{9\mu} =-1.01$,\\[-2mm]

\noindent $\blacktriangleright$ Scenario~4: $C_{9\mu}^{\rm NP} = -3 C_{9e}^{\rm NP} =-1.07$,\\[-2mm]

\noindent $\blacktriangleright$ Scenario~5: The best fit point in the six-dimensional fit given in the main article.\\[-2mm]

\begin{figure}
\includegraphics[width=8.5cm]{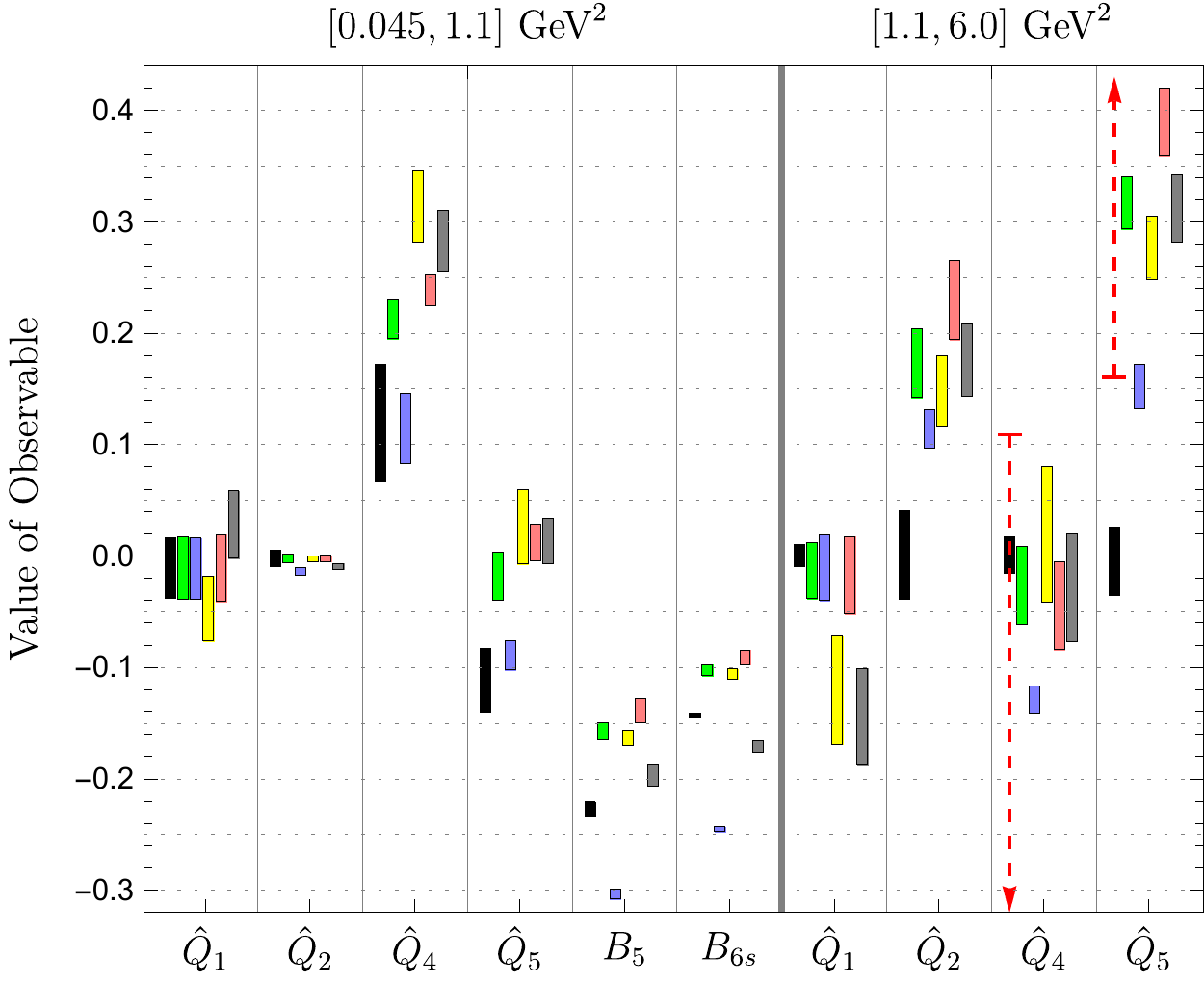}
\caption{Predictions and experimental measurements for the $\hat{Q_i}$ and $B_5$, $B_{6s}$ observables in specific bins for $B\to K^*\mu\mu$. In each case, from left to right, the predictions are given for the SM (filled black box) and for
 the Scenarios 1 to 5 (in this order) defined in Sec.~\ref{sec:oppLFUV}. The dashed red interval corresponds to the experimental measurement, when available.\label{fig:plotQi}}
\end{figure}

The results are summarised in Figs.~\ref{fig:plotR} and \ref{fig:plotQi}, where we show only the most interesting cases. We find that:
\\[-2mm]

\noindent $\blacktriangleright$ $R_K$ cannot distinguish between
Scenario 3 and the SM, but it is optimal to separate Scenarios 1 and 2 on one side and 4 and 5 on the other side, without lifting the degeneracy any further. This is true in all the three bins considered. $R_{K^\star}$ has large uncertainties at large recoil, but it has a good sensitivity to Scenario 2  in the bin [1.1,6] (although difficult to distinguish from the other NP scenarios). In the same bin $R_\phi$ fares slightly better. The low-recoil bin of $R_{K^\star}$ and $R_\phi$ is particularly promising to distinguish Scenarios 1 and 5 from each other and the SM, but only with small experimental uncertainties.\\[-2mm]

\noindent $\blacktriangleright$ $\langle \hat Q_2\rangle^{[0.045,1.1]}$ should be approximately SM-like.
It may thus be used as a control observable. \\[-2mm]

\noindent $\blacktriangleright$ The observable $\langle \hat Q_5 \rangle^{[1.1,6]}$ is able to 
discern the SM and Scenario 2 from the other four NP scenarios, depending on the experimental uncertainties. \\[-2mm]

\noindent $\blacktriangleright$ $B_5$ and $B_{6s}$ in the first bin $[0.045,1.1]$ are  sensitive to Scenario 2 and able
to distinguish it from the rest if small experimental uncertainties can be achieved.\\[-2mm]

In the near future, precise measurements of these observables will thus be instrumental in establishing the patterns for LFUV New Physics discussed in this article.


\section{Conclusions and outlook}

Over the last years, a very interesting pattern of deviations has emerged in $b\to s\ell\ell$ transitions. After the initial $P_5'$ anomaly identified in $B\to K^*\mu\mu$ by the LHCb experiment, several systematic deviations have been observed in various branching ratios. At the same time, new observables comparing electron and muon modes have been measured at LHCb ($R_K$) and Belle ($Q_{4,5}$) hinting at a violation of lepton flavour universality. A global analysis of all these deviations~\cite{Descotes-Genon:2015uva} found a preference for NP solutions with respect to the SM with high significances (below 5~$\sigma$) with distinctive features: i)  NP affect $b\to s\mu\mu$ transitions much more noticeably than $b\to see$ ones, ii) the dominant NP contribution enters the semileptonic operator ${\cal O}_{9\mu}$ and iii)  there is a strong consistency between the pattern of deviations in  $b \to s \mu\mu$ and LFUV observables.

This picture has been updated very recently due to the measurement by the LHCb experiment of a new LFUV observable, the $R_{K^*}$ ratio of $B\to K^*\ell\ell$ branching ratios, exhibiting a deviation with respect to the SM, in agreement with the expectations from earlier global analyses. This remarkable measurement calls for a reassessment of our previous analyses, which we have presented here. Considering the available data
for $b\to s\gamma$, $b\to s\mu\mu$ and $b\to see$ transitions, we performed global frequentist fits and identified one- and two-dimensional hypotheses with real NP contributions to Wilson coefficients that improve significantly the agreement between data and predictions compared to the SM, reaching significances between 5~$\sigma$ and 6~$\sigma$. 

We have also performed fits restricted to LFUV observables, showing that even this limited set of observables favours several NP hypotheses compared to the SM in a significant way and in very good agreement with 
the results from our global fit. Remarkably, a six-dimensional fit to the Wilson coefficients $\C{7(')},\C{9(')\mu},\C{10(')\mu}$ confirms the need for a large contribution to $\C{9\mu}$ and hints at contributions in $\C{9'\mu}$ and/or $\C{10\mu}$, with a SM pull reaching 5.0~$\sigma$ for the first time. We have discussed the consequences of the favoured hypotheses for models such as leptoquarks or an additional $Z'$ boson, in connection with the deviations observed in $b\to c\ell\nu$ transitions and measured by the ratios $R_{D^{(*)}}$.

On the theoretical side, hadronic uncertainties conform to theoretical expectations~\cite{Capdevila:2017ert} and unexpectedly large effects (power corrections to form factors, charm-loop contributions) are disfavoured by the significant amount of LFUV observed. However, it would be very useful to have more determinations of the form factors involved, both at low and large meson recoils, as well as refined estimates of charm-loop contributions, in order to improve the accuracy of theoretical predictions. 

In some NP models, it is possible to relate these hints of LFUV to other sectors. In the case of left-handed NP contributions explaining LFUV in both $b\to s\ell\ell$ and $b\to c\ell\nu$ transitions,
$b\to s\tau\tau$ should be enhanced by up to three orders of magnitude (thus within the reach of LHCb and Belle II) ~\cite{Crivellin:2017zlb}. On the other hand, leptoquark models explaining the same deviations yield large branching ratios (of order $10^{-5}$) for $b\to s\tau\mu$, and they provide predictions for ${\cal B}(K\to\pi\mu\mu)/{\cal B}(K\to\pi e e)$ to be measured at NA62 or KOTO~\cite{Crivellin:2016vjc}. 

Additional tests of the violation of lepton flavour universality are mandatory to determine which directions should be preferred for model building. This could be achieved through more statistics, different decay modes (such as $R_K$ with a finer binning, or $R_\phi$ for $B_s\to\phi\ell\ell$), additional observables (such as the optimised observables $Q_i$ discussed with other LFUV observables in Ref.~\cite{Capdevila:2016ivx}), and different experimental settings (such as Belle II). These measurements should prove highly instrumental in exploiting the full potential of $b\to s\ell\ell$ decays to search for New Physics and ultimately uncover its detailed pattern.

\section*{Acknowledgments}
This work received financial support from the grant FPA2014-61478-EXP [JM, SDG, BC, JV]   and from Centro de Excelencia Severo Ochoa SEV-2012-0234 [BC]; from the EU Horizon 2020 program from the grants No 690575, No 674896 and No. 692194 [SDG]. The work of A.C. is supported by an Ambizione Grant of the Swiss National Science Foundation (PZ00P2\_154834).
J.V. is funded by the Swiss National Science Foundation.


\begin{thebibliography}{99}



\bibitem{Descotes-Genon:2013wba}
S.~Descotes-Genon, J.~Matias and J.~Virto,
``Understanding the $B\to K^*\mu^+\mu^-$ Anomaly,''
Phys.\ Rev.\ D {\bf 88} (2013) 074002
[arXiv:1307.5683 [hep-ph]].


\bibitem{Descotes-Genon:2015uva}
S.~Descotes-Genon, L.~Hofer, J.~Matias and J.~Virto,
``Global analysis of $b\to s\ell\ell$ anomalies,''
JHEP {\bf 1606} (2016) 092
[arXiv:1510.04239 [hep-ph]].


\bibitem{Altmannshofer:2014rta}
W.~Altmannshofer and D.~M.~Straub,
``New physics in $b\rightarrow s$ transitions after LHC run 1,''
Eur.\ Phys.\ J.\ C {\bf 75} (2015) no.8,  382
[arXiv:1411.3161 [hep-ph]].


\bibitem{Hurth:2016fbr}
T.~Hurth, F.~Mahmoudi and S.~Neshatpour,
``On the anomalies in the latest LHCb data,''
Nucl.\ Phys.\ B {\bf 909} (2016) 737
[arXiv:1603.00865 [hep-ph]].



\bibitem{Becirevic:2015asa}
  D.~Becirevic, S.~Fajfer and N.~Kosnik,
  ``Lepton flavor nonuniversality in $b\to s\ell^+\ell^-$ processes,''
  Phys.\ Rev.\ D {\bf 92} (2015) no.1,  014016
  [arXiv:1503.09024 [hep-ph]].

\bibitem{Becirevic:2016yqi}
D.~Becirevic, S.~Fajfer, N.~Kosnik and O.~Sumensari,
``Leptoquark model to explain the $B$-physics anomalies, $R_K$ and $R_D$,''
Phys.\ Rev.\ D {\bf 94} (2016) no.11,  115021
[arXiv:1608.08501 [hep-ph]].


\bibitem{DescotesGenon:2012zf}
S.~Descotes-Genon, J.~Matias, M.~Ramon and J.~Virto,
``Implications from clean observables for the binned analysis of $B \to K*\mu^+\mu^-$ at large recoil,''
JHEP {\bf 1301} (2013) 048
[arXiv:1207.2753 [hep-ph]].


\bibitem{Blake:2017fyh}
T.~Blake, U.~Egede, P.~Owen, G.~Pomery and K.~A.~Petridis,
``An empirical model of the long-distance contributions to $\bar{B}^{0} \rightarrow \bar{K}^{*0}\mu^{+}\mu^{-}$ transitions,''
arXiv:1709.03921 [hep-ph].


\bibitem{Khodjamirian:2010vf}
A.~Khodjamirian, T.~Mannel, A.~A.~Pivovarov and Y.-M.~Wang,
``Charm-loop effect in $B \to K^{(*)} \ell^{+} \ell^{-}$ and $B\to K^*\gamma$,''
JHEP {\bf 1009} (2010) 089
[arXiv:1006.4945 [hep-ph]].


\bibitem{Aaij:2014pli}
R.~Aaij {\it et al.} [LHCb Collaboration],
``Differential branching fractions and isospin asymmetries of $B \to K^{(*)} \mu^+ \mu^-$ decays,''
JHEP {\bf 1406} (2014) 133
[arXiv:1403.8044 [hep-ex]].


\bibitem{Aaij:2016flj}
R.~Aaij {\it et al.} [LHCb Collaboration],
``Measurements of the S-wave fraction in $B^{0}\rightarrow K^{+}\pi^{-}\mu^{+}\mu^{-}$ decays and the $B^{0}\rightarrow K^{\ast}(892)^{0}\mu^{+}\mu^{-}$ differential branching fraction,''
JHEP {\bf 1611} (2016) 047
[arXiv:1606.04731 [hep-ex]].


\bibitem{Aaij:2015esa}
R.~Aaij {\it et al.} [LHCb Collaboration],
``Angular analysis and differential branching fraction of the decay $B^0_s\to\phi\mu^+\mu^-$,''
JHEP {\bf 1509} (2015) 179
[arXiv:1506.08777 [hep-ex]].


\bibitem{Descotes-Genon:2013vna}
S.~Descotes-Genon, T.~Hurth, J.~Matias and J.~Virto,
``Optimizing the basis of $B\to K^*ll$ observables in the full kinematic range,''
JHEP {\bf 1305} (2013) 137
[arXiv:1303.5794 [hep-ph]].


\bibitem{Aaij:2013qta}
R.~Aaij {\it et al.} [LHCb Collaboration],
``Measurement of Form-Factor-Independent Observables in the Decay $B^{0} \to K^{*0} \mu^+ \mu^-$,''
Phys.\ Rev.\ Lett.\  {\bf 111} (2013) 191801
[arXiv:1308.1707 [hep-ex]].


\bibitem{Aaij:2015oid}
R.~Aaij {\it et al.} [LHCb Collaboration],
``Angular analysis of the $B^{0} \to K^{*0} \mu^{+} \mu^{-}$ decay using 3 fb$^{-1}$ of integrated luminosity,''
JHEP {\bf 1602} (2016) 104
[arXiv:1512.04442 [hep-ex]].


\bibitem{Aaij:2013aln}
R.~Aaij {\it et al.} [LHCb Collaboration],
``Differential branching fraction and angular analysis of the decay $B_s^0\to\phi\mu^{+}\mu^{-}$,''
JHEP {\bf 1307} (2013) 084
[arXiv:1305.2168 [hep-ex]].


\bibitem{Abdesselam:2016llu}
A.~Abdesselam {\it et al.} [Belle Collaboration],
``Angular analysis of $B^0 \to K^\ast(892)^0 \ell^+ \ell^-$,''
arXiv:1604.04042 [hep-ex].


\bibitem{Wehle:2016yoi}
S.~Wehle {\it et al.} [Belle Collaboration],
``Lepton-Flavor-Dependent Angular Analysis of $B\to K^\ast \ell^+\ell^-$,''
Phys.\ Rev.\ Lett.\  {\bf 118} (2017) no.11,  111801
[arXiv:1612.05014 [hep-ex]].


\bibitem{Aaij:2014ora}
R.~Aaij {\it et al.} [LHCb Collaboration],
``Test of lepton universality using $B^{+}\rightarrow K^{+}\ell^{+}\ell^{-}$ decays,''
Phys.\ Rev.\ Lett.\  {\bf 113} (2014) 151601
[arXiv:1406.6482 [hep-ex]].


\bibitem{Capdevila:2016ivx}
B.~Capdevila, S.~Descotes-Genon, J.~Matias and J.~Virto,
``Assessing lepton-flavour non-universality from $B\to K^*\ell\ell$ angular analyses,''
JHEP {\bf 1610} (2016) 075
[arXiv:1605.03156 [hep-ph]].


\bibitem{ATLAS:2017dlm}
The ATLAS collaboration [ATLAS Collaboration],
``Angular analysis of $B^0_d \to K^{*}\mu^+\mu^-$ decays in $pp$ collisions at $\sqrt{s}= 8$ TeV with the ATLAS detector,''
ATLAS-CONF-2017-023.


\bibitem{CMS:2017ivg}
CMS Collaboration [CMS Collaboration],
``Measurement of the $P_1$ and $P_5'$ angular parameters of the decay $\mathrm{B}^0 \to \mathrm{K}^{*0} \mu^+ \mu^-$ in proton-proton collisions at $\sqrt{s}=8~\mathrm{TeV}$,''
CMS-PAS-BPH-15-008.


\bibitem{Altmannshofer:2017fio}
W.~Altmannshofer, C.~Niehoff, P.~Stangl and D.~M.~Straub,
``Status of the $B\rightarrow K^*\mu ^+\mu ^-$ anomaly after Moriond 2017,''
Eur.\ Phys.\ J.\ C {\bf 77} (2017) no.6,  377
[arXiv:1703.09189 [hep-ph]].


\bibitem{Aaij:2017vbb}
R.~Aaij {\it et al.} [LHCb Collaboration],
``Test of lepton universality with $B^{0} \rightarrow K^{*0}\ell^{+}\ell^{-}$ decays,''
JHEP {\bf 1708} (2017) 055
[arXiv:1705.05802 [hep-ex]].


\bibitem{Hiller:2014yaa}
G.~Hiller and M.~Schmaltz,
``$R_K$ and future $b \to s \ell \ell$ physics beyond the standard model opportunities,''
Phys.\ Rev.\ D {\bf 90} (2014) 054014
[arXiv:1408.1627 [hep-ph]].


\bibitem{Hiller:2014ula}
G.~Hiller and M.~Schmaltz,
``Diagnosing lepton-nonuniversality in $b \to s \ell \ell$,''
JHEP {\bf 1502} (2015) 055
[arXiv:1411.4773 [hep-ph]].




	
\bibitem{Grinstein:1987vj}
B.~Grinstein, R.~P.~Springer and M.~B.~Wise,
``Effective Hamiltonian for Weak Radiative B Meson Decay,''
Phys.\ Lett.\ B {\bf 202} (1988) 138.


\bibitem{Buchalla:1995vs}
G.~Buchalla, A.~J.~Buras and M.~E.~Lautenbacher,
``Weak decays beyond leading logarithms,''
Rev.\ Mod.\ Phys.\  {\bf 68} (1996) 1125
[hep-ph/9512380].


\bibitem{Amhis:2016xyh}
Y.~Amhis {\it et al.},
``Averages of $b$-hadron, $c$-hadron, and $\tau$-lepton properties as of summer 2016,''
arXiv:1612.07233 [hep-ex].


\bibitem{Aaij:2013iag}
R.~Aaij {\it et al.} [LHCb Collaboration],
``Differential branching fraction and angular analysis of the decay $B^{0} \to K^{*0} \mu^{+}\mu^{-}$,''
JHEP {\bf 1308} (2013) 131
[arXiv:1304.6325 [hep-ex]].


\bibitem{Wehle:private}
S. Wehle, private communication.

\bibitem{Khachatryan:2015isa}
V.~Khachatryan {\it et al.} [CMS Collaboration],
``Angular analysis of the decay $B^0 \to K^{*0} \mu^+ \mu^-$ from pp collisions at $\sqrt  s = 8$ TeV,''
Phys.\ Lett.\ B {\bf 753} (2016) 424
[arXiv:1507.08126 [hep-ex]].


\bibitem{Chatrchyan:2013cda}
S.~Chatrchyan {\it et al.} [CMS Collaboration],
``Angular analysis and branching fraction measurement of the decay $B^0 \to K^{*0} \mu^+\mu^-$,''
Phys.\ Lett.\ B {\bf 727} (2013) 77
[arXiv:1308.3409 [hep-ex]].


\bibitem{Descotes-Genon:2014uoa}
S.~Descotes-Genon, L.~Hofer, J.~Matias and J.~Virto,
``On the impact of power corrections in the prediction of $B \to K^*\mu^+\mu^-$ observables,''
JHEP {\bf 1412} (2014) 125
[arXiv:1407.8526 [hep-ph]].


\bibitem{Misiak:2015xwa}
M.~Misiak {\it et al.},
``Updated NNLO QCD predictions for the weak radiative B-meson decays,''
Phys.\ Rev.\ Lett.\  {\bf 114} (2015) no.22,  221801
[arXiv:1503.01789 [hep-ph]].


\bibitem{Huber:2015sra}
T.~Huber, T.~Hurth and E.~Lunghi,
``Inclusive $ \overline{B}\to {X}_s{\ell}^{+}{\ell}^{-} $ : complete angular analysis and a thorough study of collinear photons,''
JHEP {\bf 1506} (2015) 176
[arXiv:1503.04849 [hep-ph]].


\bibitem{Bobeth:2013uxa}
C.~Bobeth, M.~Gorbahn, T.~Hermann, M.~Misiak, E.~Stamou and M.~Steinhauser,
``$B_{s,d} \to \ell^+ \ell^-$ in the Standard Model with Reduced Theoretical Uncertainty,''
Phys.\ Rev.\ Lett.\  {\bf 112} (2014) 101801
[arXiv:1311.0903 [hep-ph]].


\bibitem{Straub:2015ica}
A.~Bharucha, D.~M.~Straub and R.~Zwicky,
``$B\to V\ell^+\ell^-$ in the Standard Model from light-cone sum rules,''
JHEP {\bf 1608} (2016) 098
[arXiv:1503.05534 [hep-ph]].


\bibitem{Becirevic:2017jtw}
  D.~Becirevic and O.~Sumensari,
  JHEP {\bf 1708} (2017) 104
  [arXiv:1704.05835 [hep-ph]].


\bibitem{Bordone:2016gaq}
M.~Bordone, G.~Isidori and A.~Pattori,
``On the Standard Model predictions for $R_K$ and $R_{K^*}$,''
Eur.\ Phys.\ J.\ C {\bf 76} (2016) no.8,  440
[arXiv:1605.07633 [hep-ph]].


\bibitem{Capdevila:2017ert}
B.~Capdevila, S.~Descotes-Genon, L.~Hofer and J.~Matias,
``Hadronic uncertainties in $B \to K^* \mu^+ \mu^-$: a state-of-the-art analysis,''
JHEP {\bf 1704} (2017) 016
[arXiv:1701.08672 [hep-ph]].


\bibitem{Ghosh:2014awa}
D.~Ghosh, M.~Nardecchia and S.~A.~Renner,
``Hint of Lepton Flavour Non-Universality in $B$ Meson Decays,''
JHEP {\bf 1412} (2014) 131
[arXiv:1408.4097 [hep-ph]].


\bibitem{Falkowski:2015zwa}
A.~Falkowski, M.~Nardecchia and R.~Ziegler,
``Lepton Flavor Non-Universality in B-meson Decays from a U(2) Flavor Model,''
JHEP {\bf 1511} (2015) 173
[arXiv:1509.01249 [hep-ph]].


\bibitem{Gauld:2013qba}
R.~Gauld, F.~Goertz and U.~Haisch,
``On minimal $Z'$ explanations of the $B\to K^*\mu^+\mu^-$ anomaly,''
Phys.\ Rev.\ D {\bf 89} (2014) 015005
[arXiv:1308.1959 [hep-ph]].


\bibitem{Buras:2013dea}
A.~J.~Buras, F.~De Fazio and J.~Girrbach,
``331 models facing new $b \to s\mu^+ \mu^-$ data,''
JHEP {\bf 1402} (2014) 112
[arXiv:1311.6729 [hep-ph]].


\bibitem{Jager:2017gal}
S.~J\"ager, K.~Leslie, M.~Kirk and A.~Lenz,
``Charming new physics in rare B-decays and mixing?,''
arXiv:1701.09183 [hep-ph].


\bibitem{Cox:2016epl}
P.~Cox, A.~Kusenko, O.~Sumensari and T.~T.~Yanagida,
``SU(5) Unification with TeV-scale Leptoquarks,''
JHEP {\bf 1703} (2017) 035
[arXiv:1612.03923 [hep-ph]].


\bibitem{Altmannshofer:2014cfa}
W.~Altmannshofer, S.~Gori, M.~Pospelov and I.~Yavin,
``Quark flavor transitions in $L_\mu-L_\tau$ models,''
Phys.\ Rev.\ D {\bf 89} (2014) 095033
[arXiv:1403.1269 [hep-ph]].


\bibitem{Crivellin:2015mga}
A.~Crivellin, G.~D'Ambrosio and J.~Heeck,
``Explaining $h\to\mu^\pm\tau^\mp$, $B\to K^* \mu^+\mu^-$ and $B\to K \mu^+\mu^-/B\to K e^+e^-$ in a two-Higgs-doublet model with gauged $L_\mu-L_\tau$,''
Phys.\ Rev.\ Lett.\  {\bf 114} (2015) 151801
[arXiv:1501.00993 [hep-ph]].


\bibitem{Crivellin:2015lwa}
A.~Crivellin, G.~D'Ambrosio and J.~Heeck,
``Addressing the LHC flavor anomalies with horizontal gauge symmetries,''
Phys.\ Rev.\ D {\bf 91} (2015) no.7,  075006
[arXiv:1503.03477 [hep-ph]].


\bibitem{Crivellin:2016ejn}
A.~Crivellin, J.~Fuentes-Martin, A.~Greljo and G.~Isidori,
``Lepton Flavor Non-Universality in B decays from Dynamical Yukawas,''
Phys.\ Lett.\ B {\bf 766} (2017) 77
[arXiv:1611.02703 [hep-ph]].


\bibitem{Bhatia:2017tgo}
D.~Bhatia, S.~Chakraborty and A.~Dighe,
``Neutrino mixing and $R_K$ anomaly in U(1)$_X$ models: a bottom-up approach,''
JHEP {\bf 1703} (2017) 117
[arXiv:1701.05825 [hep-ph]].


\bibitem{Belanger:2015nma}
G.~Belanger, C.~Delaunay and S.~Westhoff,
``A Dark Matter Relic From Muon Anomalies,''
Phys.\ Rev.\ D {\bf 92} (2015) 055021
[arXiv:1507.06660 [hep-ph]].


\bibitem{Boucenna:2016wpr} 
  S.~M.~Boucenna, A.~Celis, J.~Fuentes-Martin, A.~Vicente and J.~Virto,
  ``Non-abelian gauge extensions for B-decay anomalies,''
  Phys.\ Lett.\ B {\bf 760}, 214 (2016)
  [arXiv:1604.03088 [hep-ph]].


\bibitem{Boucenna:2016qad}
S.~M.~Boucenna, A.~Celis, J.~Fuentes-Martin, A.~Vicente and J.~Virto,
``Phenomenology of an $SU(2) \times SU(2) \times U(1)$ model with lepton-flavour non-universality,''
JHEP {\bf 1612} (2016) 059
[arXiv:1608.01349 [hep-ph]].


\bibitem{Gripaios:2014tna}
B.~Gripaios, M.~Nardecchia and S.~A.~Renner,
``Composite leptoquarks and anomalies in $B$-meson decays,''
JHEP {\bf 1505} (2015) 006
[arXiv:1412.1791 [hep-ph]].


\bibitem{Fajfer:2015ycq}
S.~Fajfer and N.~Kosnik,
``Vector leptoquark resolution of $R_K$ and $R_{D^{(*)}}$ puzzles,''
Phys.\ Lett.\ B {\bf 755} (2016) 270
[arXiv:1511.06024 [hep-ph]].


\bibitem{Varzielas:2015iva}
I.~de Medeiros Varzielas and G.~Hiller,
``Clues for flavor from rare lepton and quark decays,''
JHEP {\bf 1506} (2015) 072
[arXiv:1503.01084 [hep-ph]].


\bibitem{Alonso:2015sja}
R.~Alonso, B.~Grinstein and J.~Martin Camalich,
``Lepton universality violation and lepton flavor conservation in $B$-meson decays,''
JHEP {\bf 1510} (2015) 184
[arXiv:1505.05164 [hep-ph]].


\bibitem{Calibbi:2015kma}
L.~Calibbi, A.~Crivellin and T.~Ota,
``Effective Field Theory Approach to b→sℓℓ(′), B→K(*)ν$\overline{ν}$ and B→D(*)τν with Third Generation Couplings,''
Phys.\ Rev.\ Lett.\  {\bf 115} (2015) 181801
[arXiv:1506.02661 [hep-ph]].


\bibitem{Barbieri:2015yvd}
R.~Barbieri, G.~Isidori, A.~Pattori and F.~Senia,
``Anomalies in $B$-decays and $U(2)$ flavour symmetry,''
Eur.\ Phys.\ J.\ C {\bf 76} (2016) no.2,  67
[arXiv:1512.01560 [hep-ph]].


\bibitem{Sahoo:2016pet}
S.~Sahoo, R.~Mohanta and A.~K.~Giri,
``Explaining the $R_{K}$ and $R_{D^{(*)}}$ anomalies with vector leptoquarks,''
Phys.\ Rev.\ D {\bf 95} (2017) no.3,  035027
[arXiv:1609.04367 [hep-ph]].


\bibitem{Gripaios:2015gra}
B.~Gripaios, M.~Nardecchia and S.~A.~Renner,
``Linear flavour violation and anomalies in B physics,''
JHEP {\bf 1606} (2016) 083
[arXiv:1509.05020 [hep-ph]].


\bibitem{Arnan:2016cpy}
P.~Arnan, L.~Hofer, F.~Mescia and A.~Crivellin,
``Loop effects of heavy new scalars and fermions in $b\to s\mu^+\mu^-$,''
JHEP {\bf 1704} (2017) 043
[arXiv:1608.07832 [hep-ph]].


\bibitem{Mahmoudi:2014mja}
F.~Mahmoudi, S.~Neshatpour and J.~Virto,
``$B \to K^{*} \mu^{+} \mu^{-}$ optimised observables in the MSSM,''
Eur.\ Phys.\ J.\ C {\bf 74} (2014) no.6,  2927
[arXiv:1401.2145 [hep-ph]].


\bibitem{Niehoff:2015bfa}
C.~Niehoff, P.~Stangl and D.~M.~Straub,
``Violation of lepton flavour universality in composite Higgs models,''
Phys.\ Lett.\ B {\bf 747} (2015) 182
[arXiv:1503.03865 [hep-ph]].


\bibitem{Alonso:2016oyd}
R.~Alonso, B.~Grinstein and J.~Martin Camalich,
``Lifetime of $B_c^-$ Constrains Explanations for Anomalies in  $B\to D^{(*)}\tau\nu$,''
Phys.\ Rev.\ Lett.\  {\bf 118} (2017) no.8,  081802
[arXiv:1611.06676 [hep-ph]].

\bibitem{Freytsis:2015qca}
M.~Freytsis, Z.~Ligeti and J.~T.~Ruderman,
``Flavor models for $\bar{B} \to D^{(*)} \tau \bar{\nu}$,''
Phys.\ Rev.\ D {\bf 92} (2015) no.5,  054018
[arXiv:1506.08896 [hep-ph]].


\bibitem{Celis:2016azn}
A.~Celis, M.~Jung, X.~Q.~Li and A.~Pich,
``Scalar contributions to $b\to c (u) \tau \nu$ transitions,''
Phys.\ Lett.\ B {\bf 771} (2017) 168
[arXiv:1612.07757 [hep-ph]].


\bibitem{Ivanov:2017mrj}
M.~A.~Ivanov, J.~G.~Körner and C.~T.~Tran,
``Probing new physics in $\bar{B}^0 \to D^{(\ast)} \tau^- \bar\nu_{\tau}$ using the longitudinal, transverse, and normal polarization components of the tau lepton,''
Phys.\ Rev.\ D {\bf 95} (2017) no.3,  036021
[arXiv:1701.02937 [hep-ph]].


\bibitem{Faroughy:2016osc}
D.~A.~Faroughy, A.~Greljo and J.~F.~Kamenik,
``Confronting lepton flavor universality violation in B decays with high-$p_T$ tau lepton searches at LHC,''
Phys.\ Lett.\ B {\bf 764} (2017) 126
[arXiv:1609.07138 [hep-ph]].


\bibitem{Feruglio:2016gvd}
F.~Feruglio, P.~Paradisi and A.~Pattori,
``Revisiting Lepton Flavor Universality in B Decays,''
Phys.\ Rev.\ Lett.\  {\bf 118} (2017) no.1,  011801
[arXiv:1606.00524 [hep-ph]].


\bibitem{Grzadkowski:2010es}
B.~Grzadkowski, M.~Iskrzynski, M.~Misiak and J.~Rosiek,
``Dimension-Six Terms in the Standard Model Lagrangian,''
JHEP {\bf 1010} (2010) 085
[arXiv:1008.4884 [hep-ph]].


\bibitem{Celis:2017doq} 
  A.~Celis, J.~Fuentes-Martin, A.~Vicente and J.~Virto,
  ``Gauge-invariant implications of the LHCb measurements on lepton-flavor nonuniversality,''
  Phys.\ Rev.\ D {\bf 96}, no. 3, 035026 (2017)
  [arXiv:1704.05672 [hep-ph]].


\bibitem{Crivellin:2017zlb}
A.~Crivellin, D.~Müller and T.~Ota,
``Simultaneous explanation of R(D$^{(∗)}$) and b→sμ$^{+}$ μ$^{−}$: the last scalar leptoquarks standing,''
JHEP {\bf 1709} (2017) 040
[arXiv:1703.09226 [hep-ph]].


\bibitem{GarciaGarcia:2016nvr}
I.~Garcia Garcia,
``LHCb anomalies from a natural perspective,''
JHEP {\bf 1703} (2017) 040
[arXiv:1611.03507 [hep-ph]].


\bibitem{Jager:2012uw}
S.~J\"ager and J.~Martin Camalich,
``On $B \to  V \ell \ell$ at small dilepton invariant mass, power corrections, and new physics,''
JHEP {\bf 1305} (2013) 043
[arXiv:1212.2263 [hep-ph]].


\bibitem{Jager:2014rwa}
S.~J\"ager and J.~Martin Camalich,
``Reassessing the discovery potential of the $B \to K^{*} \ell^+\ell^-$ decays in the large-recoil region: SM challenges and BSM opportunities,''
Phys.\ Rev.\ D {\bf 93} (2016) no.1,  014028
[arXiv:1412.3183 [hep-ph]].


\bibitem{Ciuchini:2015qxb}
M.~Ciuchini, M.~Fedele, E.~Franco, S.~Mishima, A.~Paul, L.~Silvestrini and M.~Valli,
``$B\to K^* \ell^+ \ell^-$ decays at large recoil in the Standard Model: a theoretical reappraisal,''
JHEP {\bf 1606} (2016) 116
[arXiv:1512.07157 [hep-ph]].


\bibitem{Ciuchini:2016weo}
M.~Ciuchini, M.~Fedele, E.~Franco, S.~Mishima, A.~Paul, L.~Silvestrini and M.~Valli,
``$B\to K^*\ell^+\ell^-$ in the Standard Model: Elaborations and Interpretations,''
PoS ICHEP {\bf 2016} (2016) 584
[arXiv:1611.04338 [hep-ph]].


\bibitem{Bobeth:2017vxj} 
  C.~Bobeth, M.~Chrzaszcz, D.~van Dyk and J.~Virto,
  ``Long-distance effects in $B\to K^*\ell\ell$ from Analyticity,''
  arXiv:1707.07305 [hep-ph].


\bibitem{Aaij:2016cbx}
R.~Aaij {\it et al.} [LHCb Collaboration],
``Measurement of the phase difference between short- and long-distance amplitudes in the $B^{+}\to K^{+}\mu^{+}\mu^{-}$ decay,''
Eur.\ Phys.\ J.\ C {\bf 77} (2017) no.3,  161
[arXiv:1612.06764 [hep-ex]].


\bibitem{Crivellin:2016vjc}
A.~Crivellin, G.~D'Ambrosio, M.~Hoferichter and L.~C.~Tunstall,
``Violation of lepton flavor and lepton flavor universality in rare kaon decays,''
Phys.\ Rev.\ D {\bf 93} (2016) no.7,  074038
[arXiv:1601.00970 [hep-ph]].





\end{thebibliography}
\end{document}